\documentclass[twocolumn]{aastex62}

\usepackage{apjfonts}
\usepackage{subfigure}
\usepackage{amsmath}
\usepackage{xspace}
\usepackage{hyperref}

\newcommand{\tpop}{\texttt{twopoppy}\xspace}
\newcommand{\vfrag}{$v_{\rm frag}$\xspace}
\newcommand{\alphat}{$\alpha_{\rm t}$\xspace}
\newcommand{\tseed}{$t_{\rm seed}$\xspace}
\newcommand{\Miso}{$M_{\rm iso}$\xspace}
\newcommand{\tiso}{$t_{\rm iso}$\xspace}
\newcommand{\Mearth}{M$_\oplus$\xspace}

\providecommand{\bjdtdb}{\ensuremath{\rm {BJD_{TDB}}}}

\providecommand{\msun}{\ensuremath{\,M_\Sun}}
\providecommand{\rsun}{\ensuremath{\,R_\Sun}}
\providecommand{\lsun}{\ensuremath{\,L_\Sun}}
\providecommand{\mj}{\ensuremath{\,M_{\rm J}}}
\providecommand{\rj}{\ensuremath{\,R_{\rm J}}}

\providecommand{\re}{\ensuremath{\,R_{\rm E}}}

\shorttitle{Cold giants and inner super-earths}
\shortauthors{Chachan et al.}

\begin{document}

\title{\textbf{\large{Kepler-167e as a Probe of the Formation Histories of Cold Giants with Inner Super-Earths}}}
\correspondingauthor{Yayaati Chachan}
\email{ychachan@caltech.edu}

\author[0000-0003-1728-8269]{Yayaati Chachan}
\affil{Division of Geological and Planetary Sciences, California Institute of Technology, 1200 E California Blvd, Pasadena, CA, 91125, USA}

\author[0000-0002-4297-5506]{Paul A.\ Dalba}
\altaffiliation{NSF Astronomy and Astrophysics Postdoctoral Fellow}
\affil{Department of Earth and Planetary Sciences, University of California Riverside, 900 University Ave, Riverside, CA 92521, USA}
\affil{Department of Astronomy and Astrophysics, University of California Santa Cruz, 1156 High St., Santa Cruz, CA, USA}

\author[0000-0002-5375-4725]{Heather A. Knutson}
\affil{Division of Geological and Planetary Sciences, California Institute of Technology, 1200 E California Blvd, Pasadena, CA, 91125, USA}

\author[0000-0003-3504-5316]{Benjamin J. Fulton}
\affil{NASA Exoplanet Science Institute / Caltech-IPAC, 1200 East California Blvd, Pasadena, CA 91125, USA}

\author{Daniel Thorngren}
\affil{Institute for Research on Exoplanets (iREx), Université de Montréal, Montreal, QC, Canada}

\author{Charles Beichman}
\affil{Jet Propulsion Laboratory, California Institute of Technology, 4800 Oak Grove Drive, Pasadena, CA 91109, USA}
\affil{Caltech/IPAC-NASA Exoplanet Science Institute, 770 S. Wilson Ave, Pasadena, CA 91106, USA}

\author[0000-0002-5741-3047]{David R. Ciardi}
\affil{Caltech/IPAC-NASA Exoplanet Science Institute, 770 S. Wilson Ave, Pasadena, CA 91106, USA}

\author{Andrew W. Howard}
\affil{Cahill Center for Astrophysics, California Institute of Technology, 1216 East California Boulevard, Pasadena, CA 91125, USA}

\author{Judah Van Zandt}
\affil{Department of Physics \& Astronomy, University of California Los Angeles, Los Angeles, CA 90095, USA}

\begin{abstract}

The observed correlation between outer giant planets and inner super-Earths is emerging as an important constraint on planet formation theories. In this study we focus on Kepler-167, which is currently the only system known to contain both inner transiting super-Earths and a confirmed outer transiting gas giant companion beyond 1 au. Using long term radial velocity monitoring, we measure the mass of the gas giant Kepler-167e ($P=1071$ days) to be $1.01^{+0.16}_{-0.15}$ M$_{\rm J}$, thus confirming it as a Jupiter analog. We re-fit the \emph{Kepler} photometry to obtain updated radii for all four planets. Using a planetary structure model, we estimate that Kepler-167e contains $66\pm19$ \Mearth of solids and is significantly enriched in metals relative to its solar-metallicity host star. We use these new constraints to explore the broader question of how systems like Kepler-167 form in the pebble accretion framework for giant planet core formation. We utilize simple disk evolution models to demonstrate that more massive and metal-rich disks, which are the most favorable sites for giant planet formation, can also deliver enough solids to the inner disk to form systems of super-Earths. We use these same models to constrain the nature of Kepler-167's protoplanetary disk, and find that it likely contained $\gtrsim 300$ \Mearth of dust and was $\gtrsim 40$ au in size. These values overlap with the upper end of the observed dust mass and size distributions of Class 0 and I disks, and are also consistent with the observed occurrence rate of Jupiter analogs around sun-like stars.

\end{abstract}

\section{Introduction}

The relative rarity of Jupiter analogs around sun-like stars suggests that only $\sim 10 \%$ of protoplanetary disks provide the conditions needed for their formation \citep{Cumming2008, Wittenmyer2016, Wittenmyer2020, Fulton2021}. In contrast, close-in super-Earths and sub-Neptunes, which are $< 4$ R$_{\oplus}$ in size, appear to be much more common \citep[$30-50 \%$ occurrence rate for sun-like stars;][]{Batalha2013, Fressin2013, Petigura2018}. It was initially thought that distant gas giants and close-in super-Earths were unlikely to occur in the same system, as the growing gas giant planet was expected to prevent the formation of massive inner planets \citep{Izidoro2015, Ormel2017}. However, there is now growing observational evidence suggesting that cold gas giants are frequently accompanied by inner super-Earths \citep{Zhu2018, Bryan2019, Herman2019}. This suggests that the same protoplanetary disk properties that allow for the formation of distant giant planets are also compatible with the formation of inner super-Earths, and that the presence of an outer gas giant does not disrupt super-Earth formation. This observed correlation between inner super-Earths and outer giants therefore provides an important constraint on planet formation theories, as they must explain the formation of both types of planets in the same system.

Systems with multiple transiting super-Earths are particularly valuable for testing formation and migration models, as the transit photometry can be used to characterize their mutual inclinations and corresponding dynamical histories \citep[e.g.,][]{Masuda2020}. When combined with radial velocity (RV) or transit timing follow up to determine masses, we can additionally measure their average densities and calculate the corresponding masses in solids for these planets \citep[e.g.,][]{Dubber2019, Santerne2019, Dalba2021a}. Transit surveys like \emph{Kepler} \citep{Borucki2010} and \emph{TESS} \citep{Ricker2015} are more sensitive than RV surveys \citep[e.g.,][]{Howard2010a, Mayor2011, Rosenthal2021} to the presence of close-in planets in edge-on orbits with masses less than 10 \Mearth \citep{Winn2018}\footnote{The lower average sensitivity of RV surveys to planets below 10 \Mearth, which dominate the population of close-in planets, may explain why the correlation between super-Earth and cold Jupiter occurrence rates appears weaker in RV-only samples \citep[][Rosenthal et al. 2021]{Barbato2018}.}, making it easier to obtain a relatively complete census of the inner regions of these planetary systems. However, the probability of seeing a transit decreases with increasing semi-major axis, and the need to observe multiple transits imposes a hard limit on detectability that is a function of the duration of the survey. To date only the \emph{Kepler} survey has had the sensitivity to detect transiting planets beyond $\sim 1$ au, and they constitute a tiny fraction of the sample of known transiting planets \citep{Wang2015, Foreman-Mackey2016, Uehara2016, Kawahara2019}. It is therefore quite rare to find transiting outer companions to close-in super-Earths but that may not necessarily indicate that this configuration is rare..

Kepler-167 is unique among the sample of transiting planetary systems detected by Kepler, as it contains three close-in super-Earths accompanied by a confirmed transiting 0.9 R$_{\rm J}$ gas giant planet at 1.9 au \citep{Kipping2016, Dalba2019}. However, the measured radius of the outer gas giant is consistent with more than an order of magnitude range in its predicted mass \citep{Stevenson1982b}, making it difficult to predict its dynamical effect on the inner super-Earths. In \S\ref{sec:obs}, we present RV observations of the Kepler-167 system collected over 4 years with the HIRES instrument on the Keck telescope. In \S\ref{sec:exofast}, we carry out a joint analysis of the RV and transit data, which we use to place tight constraints on the mass and eccentricity of Kepler-167e. We also derive revised stellar properties using new \emph{Gaia} data and refit the \emph{Kepler} data for the inner super-Earths in order to provide updated radii for these planets. In \S\ref{sec:planet_prop}, we use Kepler-167e's measured mass and radius to constrain its bulk metallicity using the methods described in \cite{Thorngren2019}. Since the super-Earths are not detected in our RV data, we estimate their masses using a non-parametric mass-radius (M-R) relationship \citep{Ning2018}. This allows us to obtain an estimate of the total metal mass contained in the Kepler-167 planets and the corresponding minimum solid mass required to assemble this system.

In \S\ref{sec:formation}, we examine the implications of these results for the formation history of the Kepler-167 system. We know that the gas giant core must have formed early in order to undergo runaway accretion. In the pebble accretion paradigm, the core grows until it reaches the `isolation mass'. This mass marks the point where the core is massive enough to perturb the gas disk, forming a pressure trap beyond its orbit that effectively halts the accretion of pebbles. This pressure trap also blocks the transport of pebbles to the inner disk \citep{Morbidelli2012, Lambrechts2014}, reducing the reservoir of solids available to form super-Earths \citep{Ormel2017, Lambrechts2019}. However, pebble accretion is known to be a fairly lossy process \citep{Ormel2017ASSL, Lin2018}. That is, prior to reaching the isolation mass, a substantial amount of solids can flow past the growing giant planet core. 

We use simple dust evolution models \citep{Birnstiel2010, Birnstiel2012} to model the growth of the giant planet core in the outer disk and to track the evolution of the solid mass reservoir in the inner disk \citep{Ormel2010, Lambrechts2012}. We vary the effective pebble flux by changing key disk properties such as initial solid mass and size. This allows us to determine which disks are able to form giant planets, and to quantify the effect that the formation of the outer gas giant has on the amount of dust that reaches the inner disk. We use these models to relate the estimated solid masses of Kepler-167e and the inner super-Earths to the likely properties of its primordial disk. More broadly, we place constraints on the types of disks that can produce systems of inner super-Earths and outer gas giant companions under the pebble accretion paradigm. We summarize our conclusions in \S\ref{sec:conclusions}.

\section{Observations}\label{sec:obs}

\subsection{Archival Photometry}\label{sec:kepler}

Archival photometry of Kepler-167 exists from the \emph{Kepler} mission \citep[e.g.,][]{Borucki2010}, the \emph{Spitzer} spacecraft \citep{Dalba2019}, and the \emph{Transiting Exoplanet Survey Satellite} mission \citep[\emph{TESS};][]{Ricker2015}. The transiting planets in this system were initially discovered in the \emph{Kepler} data \citep{Kipping2016}. The \emph{Spitzer} observations specifically targeted a transit of Kepler-167e, but they only spanned part of the transit light curve. While these observations did not improve the precision of Kepler-167e's transit ephemeris, they did establish that the transit occurred at the expected time. This discovery significantly mitigated some of the uncertainty inherent to long-period exoplanets with only two observed transits, where the existence of transit timing variations (TTVs) can significantly bias initial estimates of the orbital period \citep[e.g.,][]{Dalba2016, Santerne2019}. Kepler-167 was also observed by \emph{TESS} in Cycle 2 of its primary mission and Cycle 4 of its extended mission. 

Our analysis of Kepler-167 archival photometry only uses the \emph{Kepler} data. The \emph{Spitzer} observations mitigate uncertainties in the orbital period due to possible TTVs, but do not improve the precision of the outer planet's ephemeris when we assume a constant ephemeris. The epochs of the \emph{TESS} photometry span transits of the inner planets but not the outer one. However, \emph{TESS} was designed to survey stars much brighter than Kepler-167 ($V\approx14$) and the \emph{TESS} observations are too imprecise to improve our constraints on the physical properties of the small inner super-Earths in this system.

The \emph{Kepler} spacecraft observed Kepler-167 during 17 quarters of its primary mission (May 2009 through May 2013). Observations in Quarters 1--8 were long cadence (30~minutes) while those in Quarters 9--17 were short cadence (1~minute). This observation window spanned dozens to hundreds of transits of the three inner planets and two transits of the outer giant planet \citep{Kipping2016}. We accessed the Pre-search Data Conditioning Simple Aperture Photometry \citep[PDCSAP;][]{Jenkins2010, Smith2012, Stumpe2012} through the Mikulski Archive for Space Telescopes\footnote{\url{https://archive.stsci.edu/}.} using the \texttt{lightkurve}\footnote{\url{https://docs.lightkurve.org/}} package \citep{lightkurve2018}. Although the PDCSAP data are corrected for many sources of systematic noise, we noticed a quasi-periodic variability signal in the corrected photometry for this target that is likely due to stellar rotation. We modeled this signal using Gaussian Process (GP) regression as implemented in the \texttt{celerite2}\footnote{\url{https://celerite2.readthedocs.io/en/latest/}} package built into the \texttt{exoplanet}\footnote{\url{https://docs.exoplanet.codes}} toolkit \citep{ForemanMackey2017, Luger2018, ForemanMackey2018, Agol2020, ForemanMackey2021}. We fit the long and short cadence data with quasi-periodic kernels of different widths but the same period. We determined the \emph{maximum a posteriori} parameters for the GP (see Section~\ref{sec:modelling}) with a numerical optimization method \citep{pymc3}. Then, we subtracted the GP signal from the long and short cadence data before fitting for the transits. 

\subsection{Keck-HIRES Spectroscopy}\label{sec:hires}

We obtained spectroscopic observations of Kepler-167 using the high-resolution echelle spectrometer \citep[HIRES;][]{Vogt1994} on the Keck~I telescope at the W. M. Keck Observatory. We first processed a moderate signal-to-noise (S/N) ratio (40) reconnaissance spectrum with \texttt{SpecMatch}\footnote{\url{https://github.com/petigura/specmatch-syn}} \citep{Petigura2015,Petigura2017b}. This spectrum was originally acquired for the Kepler-167 discovery effort \citep{Kipping2016} and processed with the Stellar Parameter Classification pipeline \citep[SPC;][]{Buchhave2012}. The stellar properties that we obtain from \texttt{SpecMatch} are in close agreement with those published by \citet{Kipping2016}. Specifically, the stellar metallicity ([Fe/H]), effective temperature ($T_{\rm eff}$), and projected rotational velocity ($v\sin{i}$) are $-0.02\pm0.09$~dex, $4830\pm100$~K, and $2.0\pm1.0$~km~s$^{-1}$, respectively. 

We acquired 13 additional spectra between 2017 August 23 and 2020 December 4 with S/N ranging between 40--52. In these observations, the starlight was passed through a heated iodine cell in order to allow us to obtain a more precise wavelength calibration. Owing to the faintness of Kepler-167 ($V\approx14$), we did not utilize a high S/N template spectrum for our radial velocity analysis, and instead substituted a best-match template from another star (HD~16160, $T_{\rm eff} = 4720 \pm 110$~K, log~$g = 4.57 \pm 0.10$, [Fe/H]~$= -0.02 \pm 0.09$) in the HIRES spectral library \citep{Yee2017, Dalba2020}. Aside from the template substitution, the data reduction and Doppler analysis followed the standard procedures of the California Planet Search \citep{Howard2010, Howard2016}. This analysis produced RV measurements of Kepler-167 with a median internal precision of 3--5~m~s$^{-1}$. We then added an additional 6.2~m~s$^{-1}$ error in quadrature, which is a conservative estimate of the average uncertainty incurred by the match--template technique \citep{Dalba2020}. The resulting individual RV measurements are listed in Table~\ref{tab:rv}. We also include the corresponding $S_{\rm HK}$ activity indicators derived from the Ca II H and K spectral lines \citep{Wright2004,Isaacson2010}. We see no evidence for any covariance between the measured radial velocities and this activity indicator.

\begin{deluxetable}{ccc}
\tablecaption{RV Measurements of Kepler-167. \label{tab:rv}}
\tablehead{
  \colhead{BJD$_{\rm TDB}$} & 
  \colhead{RV (m s$^{-1}$)} &
  \colhead{$S_{\rm HK}$}}
  \startdata
    2457988.955812 &   6.1$\pm$7.2 & 0.195$\pm$0.001 \\
    2458300.972784 &  16.1$\pm$7.0 & 0.248$\pm$0.001 \\
    2458328.854650 &  22.6$\pm$7.0 & 0.235$\pm$0.001 \\
    2458363.864254 &   6.3$\pm$7.2 & 0.267$\pm$0.001 \\
    2458385.796973 &  18.1$\pm$7.2 & 0.149$\pm$0.001 \\
    2458645.987639 & -11.8$\pm$7.3 & 0.239$\pm$0.001 \\
    2458662.992470 & -26.3$\pm$7.0 & 0.236$\pm$0.001 \\
    2458710.986137 & -24.2$\pm$8.0 & 0.167$\pm$0.001 \\
    2458797.818538 & -23.2$\pm$8.1 & 0.148$\pm$0.001 \\
    2459072.003261 &   5.6$\pm$7.3 & 0.168$\pm$0.001 \\
    2459101.942839 &   2.1$\pm$7.2 & 0.166$\pm$0.001 \\
    2459118.833084 &  14.4$\pm$7.1 & 0.186$\pm$0.001 \\
    2459187.714084 &  29.9$\pm$7.3 & 0.247$\pm$0.001 \\
  \enddata
  (This table is available in machine-readable format.)
\end{deluxetable}

\subsection{Constraints on companion properties}
High resolution images from UK Infrared Telescope Survey and Keck NIRC2 reveal the presence of a companion 2" to the NE of Kepler-167 \citep{Kipping2016}. This companion is $\sim 5$ magnitudes fainter than Kepler-167 in the \emph{Kepler} bandpass. \cite{Kipping2016} could not establish whether this companion is bound to Kepler-167. However, the parallax ($\Delta \varpi$ = $0.15 \pm 0.22$ mas) and proper motion ($\Delta$ $\mu$\_ra = $0.43 \pm 0.27$ mas yr$^{-1}$, $\Delta$ $\mu$\_dec = $1.2 \pm 0.3$ mas yr$^{-1}$) measurements from \emph{Gaia} DR3 for these two sources agree to within the uncertainties \citep{Gaia2021}, thus indicating that they are co-moving. The companion has a spectral type of M4V and is $\sim 0.2$ \msun, which is consistent with the J$-$K color measured in \cite{Kipping2016}. Given the large separation of 2" ($\sim 700$ au, $P > 15,000$ yrs) and the small stellar mass, the companion's effect on the RVs is negligible. Additionally, although the companion's light contaminates the \emph{Kepler} light curves, its effect on the measured planetary radii is insignificant. The companion is 100 times fainter than Kepler-167 in the \emph{Kepler} bandpass and thus the true radii of Kepler-167's planets are only $\sqrt{1.01 / 1} = 1.005$ larger than our measurements \citep{Ciardi2015}. This effect is an order of magnitude smaller than our uncertainties on the radii.

\section{Model Fitting and Parameter Estimation}\label{sec:exofast}

\begin{deluxetable}{lcc}
\tabletypesize{\scriptsize}
\tablecaption{Median values and 68\% confidence intervals for the stellar parameters for Kepler-167. \label{tab:star}}
\tablehead{\colhead{~~~Parameter} & \colhead{Description} & \colhead{Values}}
\startdata
\multicolumn{2}{l}{Informative Priors:}& \smallskip\\
~~~~$T_{\rm eff}$\dotfill &Effective Temperature (K)\dotfill & $\mathcal{N}(4830,100)$\\
~~~~$[{\rm Fe/H}]$\dotfill &Metallicity (dex)\dotfill & $\mathcal{N}(-0.02,0.09)$\\
~~~~$\varpi$\dotfill &Parallax (mas)\dotfill & $\mathcal{N}(2.944,0.018)$\\
~~~~$A_V$\dotfill &V-band extinction (mag)\dotfill & $\mathcal{U}(0,0.4204)$\\
\smallskip\\\multicolumn{2}{l}{Stellar Parameters from SED-only fit:}&\smallskip\\
~~~~$M_*$\dotfill &Mass (\msun)\dotfill &$0.777^{+0.034}_{-0.031}$\\
~~~~$R_*$\dotfill &Radius (\rsun)\dotfill &$0.749\pm0.020$\\
~~~~$L_*$\dotfill &Luminosity (\lsun)\dotfill &$0.289^{+0.017}_{-0.020}$\\
~~~~$F_{Bol}$\dotfill &Bolometric Flux (cgs)\dotfill &$8.02\times10^{-11}$$^{+4.7\times10^{-12}}_{-5.7\times10^{-12}}$\\
~~~~$\rho_*$\dotfill &Density (cgs)\dotfill &$2.60^{+0.23}_{-0.20}$\\
~~~~$\log{g}$\dotfill &Surface gravity (cgs)\dotfill &$4.579^{+0.027}_{-0.025}$\\
~~~~$T_{\rm eff}$\dotfill &Effective Temperature (K)\dotfill &$4884^{+69}_{-75}$\\
~~~~$[{\rm Fe/H}]$\dotfill &Metallicity (dex)\dotfill &$0.020\pm0.067$\\
~~~~$[{\rm Fe/H}]_{0}$\dotfill &Initial Metallicity$^{2}$ \dotfill &$0.024^{+0.069}_{-0.067}$\\
~~~~$Age$\dotfill &Age (Gyr)\dotfill &$7.1^{+4.4}_{-4.6}$\\
~~~~$EEP$\dotfill &Equal Evolutionary Phase$^{3}$ \dotfill &$339^{+12}_{-28}$\\
~~~~$A_V$\dotfill &V-band extinction (mag)\dotfill &$0.277^{+0.098}_{-0.13}$\\
~~~~$\sigma_{SED}$\dotfill &SED photometry error scaling \dotfill &$1.23^{+0.48}_{-0.32}$\\
~~~~$\varpi$\dotfill &Parallax (mas)\dotfill &$2.945\pm0.018$\\
~~~~$d$\dotfill &Distance (pc)\dotfill &$339.6\pm2.1$\\
\enddata
\tablenotetext{}{See Table~3 in \citet{Eastman2019} for a detailed description of all parameters and all default (non-informative) priors beyond those specified here. $\mathcal{N}(a,b)$ denotes a normal distribution with mean $a$ and variance $b^2$. $\mathcal{U}(a,b)$ denotes a uniform distribution over the interval [$a$,$b$].}
\tablenotetext{a}{Initial metallicity is that of the star when it formed.}
\tablenotetext{b}{Corresponds to static points in a star's evolutionary history. See Section~2 of \citet{Dotter2016}.}
\end{deluxetable}
 
The combined \emph{Kepler} data set for Kepler-167 system contains over 100,000 individual data points measuring signals from four separate transiting planets. When combined with the complexity of a model combining stellar, RV, and transit data, this data volume makes it computationally intractable to fit a single joint model. Instead, we separated the modelling of this system into three parts: the stellar parameters, the long period gas giant Kepler-167e, and the three inner super-Earths.

\begin{figure*}
    \centering
    \includegraphics[width=\textwidth]{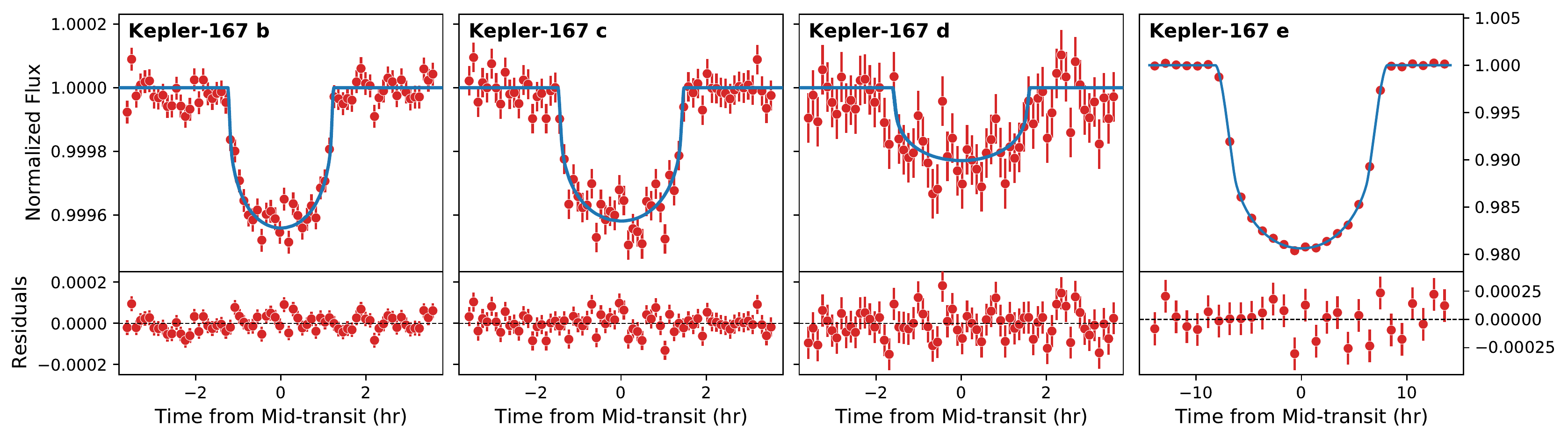}
    \caption{Optical \emph{Kepler} transit photometry of the four Kepler-167 planets folded on their best fit ephemeris along with their best fit models (blue). Note the difference in scale in the panel for Kepler-167e. The data (red) are shown in bins of 10~minutes for the inner planets and 1~hour for the outer planet, although we note that the models were fit to the unbinned data.}
    \label{fig:transits}
\end{figure*}

\begin{deluxetable*}{lccccc}
\tabletypesize{\scriptsize}
\tablecaption{Median values and 68\% confidence interval of the parameters for each of the Kepler-167 planets. \label{tab:planet}}
\tablehead{\colhead{~~~Parameter} & \colhead{Description} & \multicolumn{4}{c}{Values}}
\startdata
\smallskip\\\multicolumn{2}{l}{Planetary Parameters:}&b&c&d&e\smallskip\\
~~~~$P$\dotfill &Period (days)\dotfill &$4.3931539^{+0.0000048}_{-0.0000046}$&$7.406106\pm0.000010$&$21.80379^{+0.00013}_{-0.00018}$&$1071.23205^{+0.00059}_{-0.00058}$\\
~~~~$T_C$\dotfill &Time of conjunction (\bjdtdb)\dotfill &$2455831.78065\pm0.00039$&$2455552.15797^{+0.00070}_{-0.00071}$&$2455669.7888^{+0.0028}_{-0.0023}$&$2455253.28756\pm0.00039$\\
~~~~$R_P$\dotfill &Radius (\re)\dotfill &$1.718\pm0.070$&$1.674\pm0.069$&$1.238\pm0.064$&$10.16\pm0.42$\\
~~~~$a$\dotfill &Semi-major axis (au)\dotfill &$0.04825\pm0.00070$&$0.0684\pm0.0010$&$0.1404\pm0.0020$&$1.883\pm0.027$\\
~~~~$i$\dotfill &Inclination (Degrees)\dotfill &$88.3^{+1.6}_{-1.2}$&$88.48^{+0.88}_{-1.0}$&$89.26\pm0.50$&$89.9720^{+0.0069}_{-0.0079}$\\
~~~~$b$\dotfill &Transit impact parameter \dotfill &$0.41^{+0.38}_{-0.28}$&$0.52^{+0.30}_{-0.35}$&$0.52\pm0.35$&$0.271^{+0.051}_{-0.073}$\\
~~~~Depth\dotfill &Transit depth \dotfill &$0.0004407\pm0.0000078$&$0.0004187^{+0.0000090}_{-0.0000089}$&$0.000229\pm0.000015$&$0.01540^{+0.00027}_{-0.00032}$\\
~~~~$T_{eq}$\dotfill &Equilibrium temperature$^{a}$ (K)\dotfill &$918\pm27$&$771\pm23$&$538\pm16$&$134.4\pm4.0$\\
~~~~$e$\dotfill &Eccentricity$^{b}$ \dotfill &0&0&0&$<0.29$\\
~~~~$\omega_*$\dotfill &Argument of Periastron (rad)\dotfill &--&--&--&$-0.4^{+2.0}_{-1.0}$\\
~~~~$K$\dotfill &RV semi-amplitude (m s$^{-1}$)\dotfill &--&--&--&$23.7^{+3.7}_{-3.5}$\\
~~~~$M_P$\dotfill &Mass (\mj)\dotfill &--&--&--&$1.01^{+0.16}_{-0.15}$\\
~~~~$\rho_P$\dotfill &Density (g cm$^{-3}$)\dotfill &--&--&--&$1.68^{+0.34}_{-0.33}$\\
~~~~$\dot{\gamma}$\dotfill & RV slope (m s$^{-1}$ day$^{-1}$)\dotfill & \multicolumn{4}{c}{$-0.0049^{+0.0068}_{-0.0066}$}\\
\smallskip\\\multicolumn{2}{l}{Kepler Parameters:}& \smallskip\\
~~~~$u_{1}$\dotfill &Linear limb-darkening coefficient \dotfill &\multicolumn{4}{c}{$0.616^{+0.036}_{-0.034}$}\\
~~~~$u_{2}$\dotfill &Quadratic limb-darkening coefficient \dotfill &\multicolumn{4}{c}{$0.126\pm0.096$}\\
\smallskip\\\multicolumn{2}{l}{Keck-HIRES Parameters:}& \smallskip\\
~~~~$\gamma_{\rm rel}$\dotfill &Relative RV Offset (m s$^{-1}$)\dotfill &\multicolumn{4}{c}{$-2.49^{+0.95}_{-0.92}$}\\
~~~~$\sigma_J$\dotfill &RV Jitter (m s$^{-1}$)\dotfill &\multicolumn{4}{c}{$3.1^{+2.5}_{-1.6}$}\\
\enddata
\tablenotetext{a}{Assumes a Bond albedo of 0.3 for Kepler-167e and 0 for all other planets. Assumes perfect heat redistribution.}
\tablenotetext{b}{Fixed to 0 for all planets except Kepler-167e, for which we report the $3\sigma$ upper limit.}
\end{deluxetable*}

We first determined the stellar parameters by fitting archival photometry of Kepler-167 from the \emph{Gaia} \citep{Gaia2018}, 2MASS \citep{Cutri2003}, and WISE \citep{Cutri2014} surveys with a model spectral energy distribution to constrain the stellar properties. We employed the \texttt{EXOFASTv2} modelling suite \citep{Eastman2013, Eastman2019} to conduct this fit. The fit included the MESA Isochrones and Stellar Tracks (MIST) stellar evolution models \citep{Paxton2011, Paxton2013, Paxton2015, Choi2016, Dotter2016}, which provided constraints on the stellar mass and age. We placed normal priors on [Fe/H] ($-0.02\pm0.09$) and $T_{\rm eff}$ ($4830\pm100$~K) from the \texttt{SpecMatch} analysis of the HIRES iodine-free spectrum (Section~\ref{sec:hires}) and an upper limit on the line-of-sight extinction ($A_V<0.4204$) from galactic reddening maps \citep{Schlafly2011}. The parallax of Kepler-167 as measured by \emph{Gaia} Early Data Release 3 \citep{Gaia2020} and corrected according to \citet{Lindegren2020} is $2.944\pm0.018$~mas, which we applied as a normal prior in the fit. We also enforced a noise floor of 2\% on the bolometric flux as suggested by \citet{Tayar2020}. We checked that this fit met the default criteria for convergence in \texttt{EXOFASTv2}, which includes at least 1,000 independent draws from the posterior and a Gelman--Rubin statistic below 1.01 for each parameter. The resulting stellar parameters and their corresponding priors are summarized in Table~\ref{tab:star}. All of our stellar parameters are consistent with those derived by \citet{Kipping2016} to the 2$\sigma$ level. We inflated the widths of the uncertainties on the stellar mass and radius priors to 5\% and 4\%, respectively, prior to using these values to calculate absolute planetary parameters. This inflation accounts for systematic uncertainty floors set by imperfect models of stellar evolution \citet{Tayar2020}.

\begin{figure}
    \centering
    \includegraphics[width=\columnwidth]{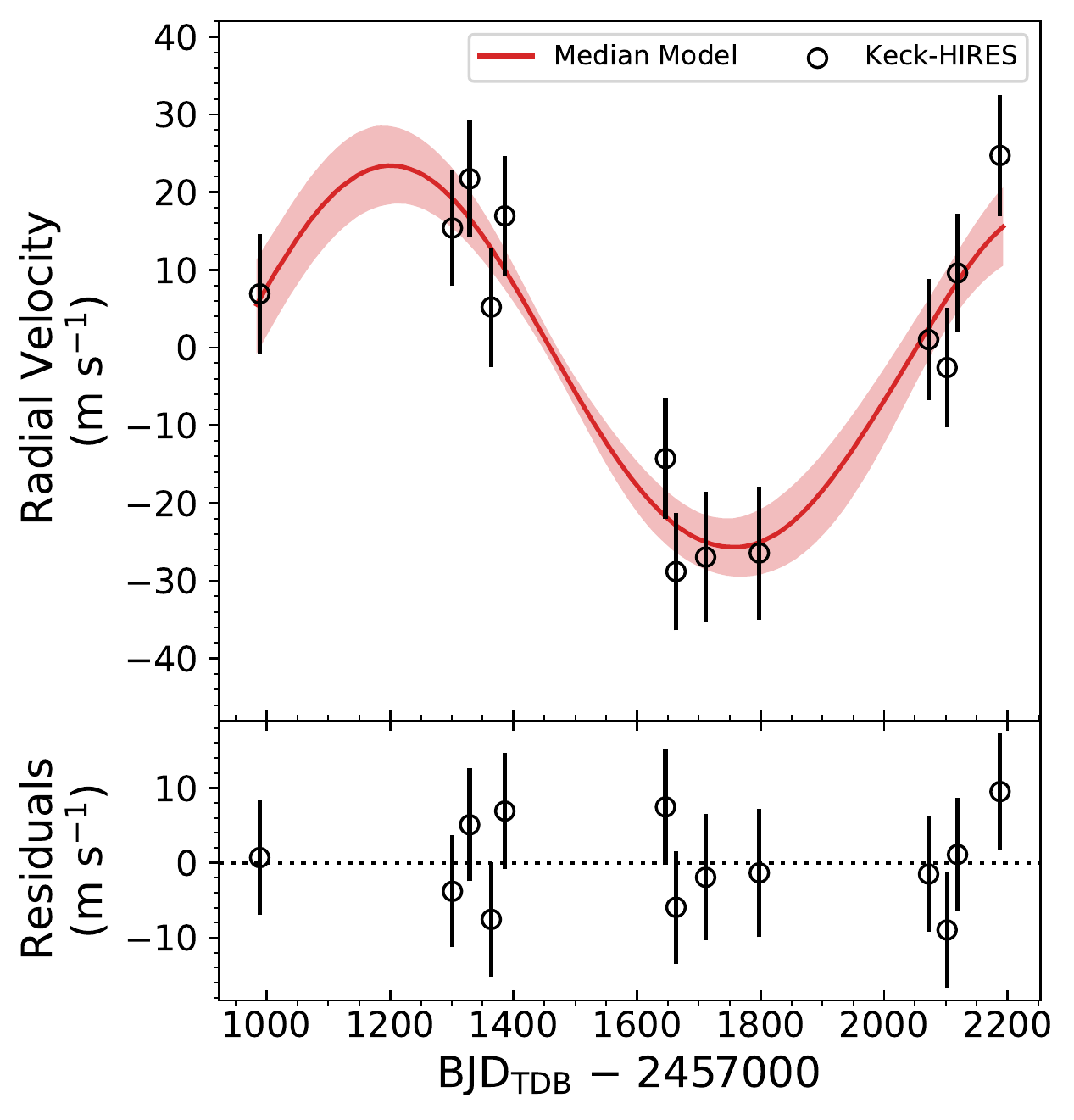}
    \caption{Keck-HIRES RV time series data with the median model for Kepler-167e and the corresponding 68\% credible interval overplotted.}
    \label{fig:rv}
\end{figure}

Second, we conducted a joint fit of the Keck-HIRES RVs and the two \emph{Kepler} transits of Kepler-167e using \texttt{exoplanet} \citep{ForemanMackey2021}. We allowed for orbital eccentricity and a long-term acceleration (slope) in the RV. For the long-cadence photometry, we numerically integrated the transit model over the appropriate time bin in order to account for the effect of these longer integrations on the shape of the transit light curve. We checked for convergence using both the effective sample size and the Gelman--Rubin statistic, which we required to be greater than 1,000 and less than 1.01 for all parameters, respectively. The resulting parameters and models are provided in Table~\ref{tab:planet} and Figures~\ref{fig:transits} and \ref{fig:rv}.

Finally, we conducted a separate fit to the transit photometry of the inner three planets in the Kepler-167 system. In order to simplify the fit and reduce the convergence time, we fixed the orbital eccentricity of these planets to zero. This is likely a valid assumption for Kepler-167 b and c, which both orbit within 0.1~au and have either been tidally circularized or have a sufficiently low eccentricity that the effect on the transit shape is negligible \citep[e.g.,][]{Mills2019}. However, Kepler-167d is far enough from these two planets to maintain some orbital eccentricity. Using the same \emph{Kepler} data set, \citet{Kipping2016} demonstrated an upper limit on eccentricity of 0.12. This indicates that there is no detectable deviation from the expected transit shape for a circular orbit in the \emph{Kepler} photometry, and we therefore should not introduce any additional error by fixing this planet's eccentricity to zero in our fits. As in the previous fit, we numerically integrated the model to account for the integration time when fitting the long cadence data. 
We applied the priors on limb darkening parameters from our fit to the Kepler-167e photometry, which has a much higher SNR than that of the inner super-Earths. We again gauged convergence with the effective sample size and the Gelman-Rubin statistic, for which we achieved $>$1,000 and $<$1.01 for all parameters. The best fit models are shown along with the transit and RV data in Figures \ref{fig:transits} and \ref{fig:rv}, respectively. The resulting planetary parameters are listed in Table~\ref{tab:planet}. All of the shared planetary parameters between our work and that of \citet{Kipping2016} are consistent at the 2$\sigma$ level. We note that the median values of all of the planetary radii are slightly larger than those from \citet{Kipping2016} owing to the increase in stellar radii derived from the updated \emph{Gaia} parallax.

\section{A closer look at the Kepler-167 planets}
\label{sec:planet_prop}

\subsection{Kepler-167e in the context of other cold giants}\label{sec:k167_bulk_metal}

\begin{figure}
    \centering
    \includegraphics[width=\linewidth]{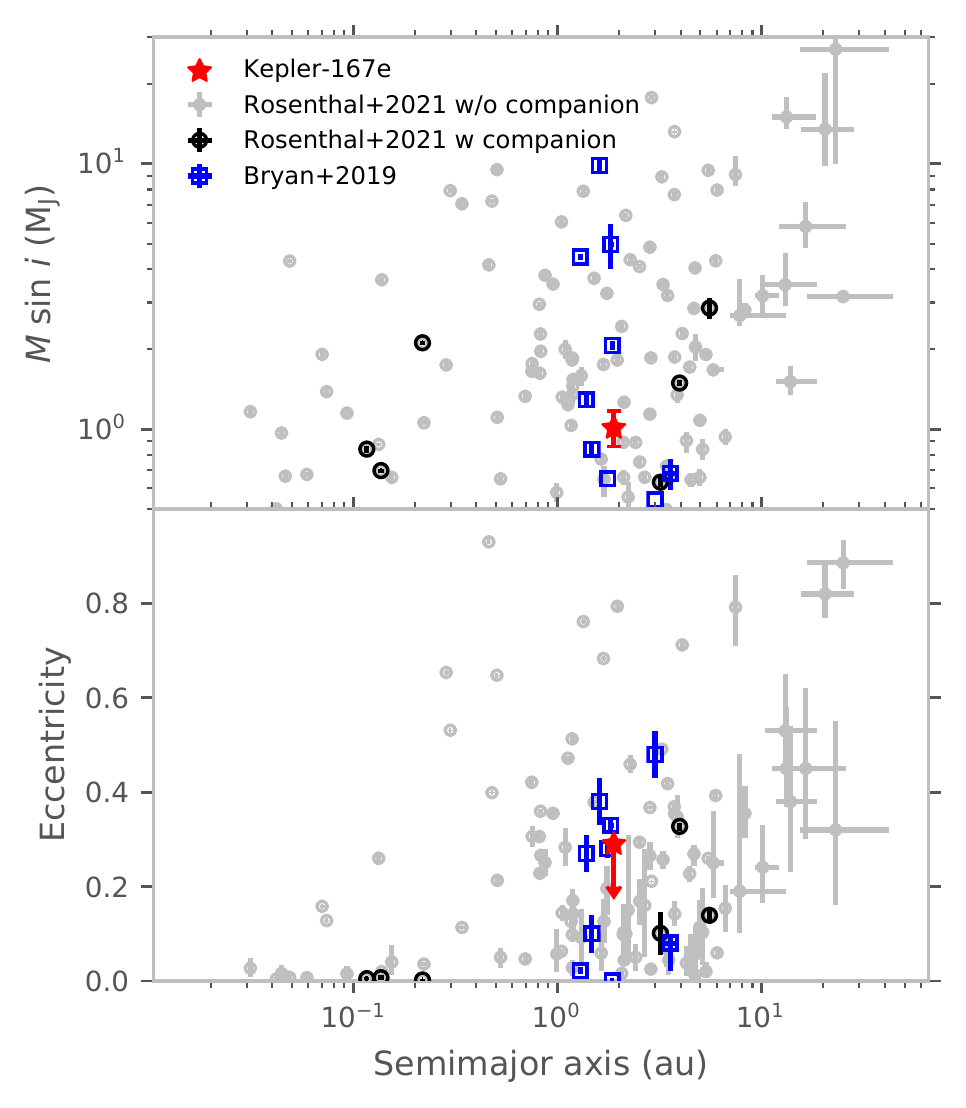}
    \caption{Planet mass and eccentricity vs semimajor axis for planets more massive than 0.5 M$_{\rm J}$ in the California Legacy Survey sample \citep{Rosenthal2021}. Giant planets in this sample are shown in two different colors, depending on whether or not they have a detectable inner companion less than 0.1 M$_{\rm J}$. We also plot resolved giant planet companions of super-Earths systems studied in \cite{Bryan2019}. Kepler-167e is marked in red ($3 \sigma$ upper limit for eccentricity) and has properties that are typical of long period giant planets.}
    \label{fig:K167e_context}
\end{figure}

Our new radial velocity observations allow us to detect the radial velocity signal from Kepler-167e with high statistical significance (6$\sigma$). We find a measured mass of $1.01\pm0.16$ \mj~and a $3 \sigma$ upper limit of $0.29$ for the orbital eccentricity. This new mass measurement for Kepler-167e allows us to place it in the context of the broader population of long-period gas giant planets ($> 0.5$ M$_{\rm J}$) from radial velocity surveys. We focus our comparison on the sample of planets detected in the California Legacy Survey \citep[CLS,][]{Rosenthal2021}, shown in Figure~\ref{fig:K167e_context}. Although there have been several other long-term radial velocity surveys capable of detecting Jupiter analogs \citep[e.g.,][]{Rowan2016, Wittenmyer2016, Wittenmyer2020}, the CLS includes data sets for more than seven hundred stars with baselines spanning close to three decades, making it one of the largest and most complete surveys for Jupiter analogs published to date. This allows us to obtain a (relatively) unbiased sample of long period gas giant planets both with and without inner super-Earth companions. We mark giant planets from the \cite{Rosenthal2021} sample that harbor an inner companion ($< 0.1$ M$_{\rm J}$) in black. We additionally supplement this sample with the set of (RV) giant planet companions to transiting super-Earths with resolved orbits in \cite{Bryan2019}. Although \cite{Bryan2019} had much better sensitivity to small inner companions compared to CLS, the systems in their sample have shorter RV baselines and the companion orbits are therefore biased towards shorter orbital periods than in the CLS sample.

We find that the mass and eccentricity of Kepler-167e are fairly typical of other long period giant planets, both with and without detected inner super-Earths. Despite the faintness of the host star and the relative sparseness of our radial velocity data, our joint fit with transit data results in constraints on mass and eccentricity that are comparable in precision to those of the non-transiting planets detected in these RV surveys. We find that the sample of radial velocity planets span a wide range of masses at Kepler-167e's location, although these data are relatively insensitive to planets smaller than $\sim0.5$ \mj~ \citep{Rosenthal2021} at these large separations. While Kepler-167e's mass may be typical of planets at these separations, it appears to have a relatively low orbital eccentricity. Although it is possible that a higher orbital eccentricity might have destabilized the system of inner super-Earths, we note that there are multiple examples of eccentric gas giants in the \cite{Bryan2019} sample with inner transiting super-Earths.

\begin{figure}
    \centering
    \includegraphics[width=\linewidth]{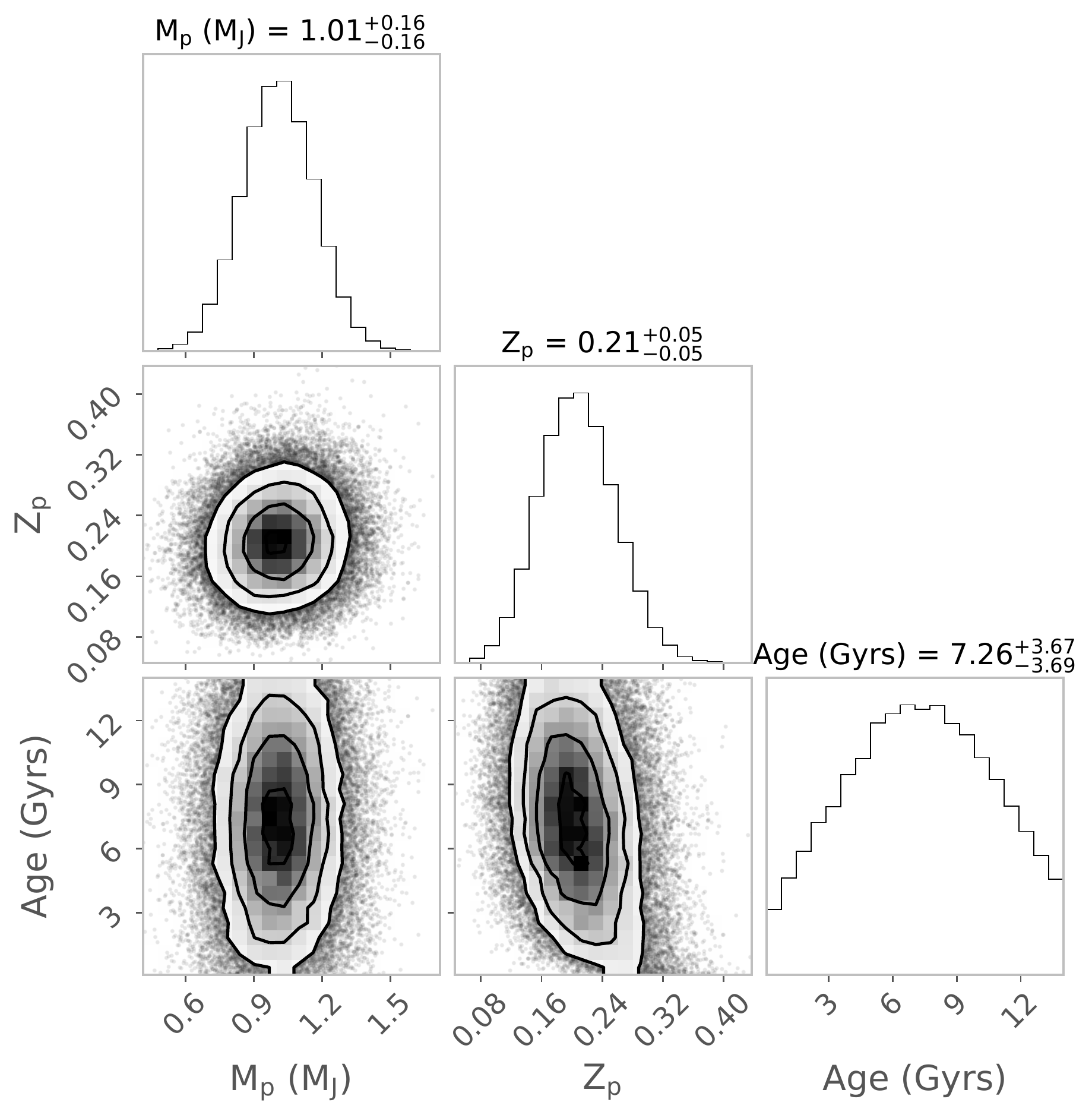}
    \caption{Posterior probability distribution for bulk metallicity and its covariance with the observational priors on planetary mass and age. Kepler-167e's radius, mass, and age are used to infer its bulk metal content from planetary evolution modeling as described in \cite{Thorngren2016} and \cite{Thorngren2019}.}
    \label{fig:K167e_Z}
\end{figure}

The fact that Kepler-167e also transits its host star provides us with a unique opportunity to use its measured mass, radius, and age to constrain its bulk metallicity and absolute metal content. Since Kepler-167e has a mass that is indistinguishable from that of Jupiter but is 10\% smaller than Jupiter in radius, we can immediately surmise that the planet has a higher bulk metal content. We use the giant planet interior structure and evolution model described in \cite{Thorngren2016} and \cite{Thorngren2019} to quantify the corresponding bulk metallicity for Kepler-167e. In this approach, a Bayesian statistical model is used to infer a planet's bulk metallicity from its mass, age, and radius. The planetary mass and age observations are used as priors and the planetary radius is the independent measurement that yields an estimate of the bulk metallicity. For more details on the giant planet interior structure model and the Bayesian method used to make these inferences, we refer the reader to \cite{Thorngren2019}. The only difference between the models described there and the ones used here is that we updated the equation of state for hydrogen and helium to the one given in \cite{Chabrier2019}. Figure~\ref{fig:K167e_Z} shows the resulting posterior probability distribution for bulk metallicity and its covariance with the observational priors on planetary mass and age.

With a bulk metallicity of $0.21 \pm 0.05$, Kepler-167e is significantly enriched in metals relative to its host star, which has a metallicity [Fe/H] of $0.02 \pm 0.07$ (bulk metallicity $Z_* = 0.015 \pm 0.003$ assuming solar $Z = 0.014$, \citealp{Asplund2009}). A bulk metallicity of $0.21 \pm 0.05$ and planet mass of $1.01^{+0.16}_{-0.15}$ M$_{\rm J}$ translates to an absolute metal content of $66^{+20}_{-18}$ \Mearth. Kepler-167e has a bulk metallicity that is fairly typical for gas giants planets in this mass range: transiting planets with measured bulk metallicities in the mass range $0.85-1.17$ M$_{\rm J}$ have a median bulk metallicity of 0.22 \citep{Thorngren2019}. Although we cannot tell how the metals are partitioned between the core and the envelope, the total metal content of the planet nonetheless provides a useful constraint on the solid inventory of the disk that Kepler-167e formed in. We discuss the implications of this measurement in more detail in \S\ref{sec:modelling}.

To date, only four other transiting giant planets with orbital periods $> 200 $ days have precisely measured masses: the circumbinary planets Kepler-16b \citep{Doyle2011} and Kepler-34b \citep{Welsh2012}, Kepler-1514b \citep{Dalba2020}, and Kepler-1704b, \citep{Dalba2021b}. Amongst the planets around single stars, Kepler-1704b is a massive giant (4.15~\mj) on a highly eccentric ($e = 0.92$) orbit and therefore unlikely to harbor any undetected inner companions. Kepler-1514b is also quite massive (5.3~\mj) and has a high eccentricity ($e = 0.4$) but it is accompanied by a single transiting inner super-Earth. Kepler-167e is only planet in this sample that has multiple transiting inner super-Earths. Both Kepler-1514b ($1.11 \pm 0.02$~\rj) and Kepler-1704b ($1.07 \pm 0.04$~\rj) are larger than Jupiter in size and therefore their bulk metallicities ($0.06^{+0.03}_{-0.02}$ and $0.12 \pm 0.04$ respectively) are lower than Kepler-167e's. However, given their large masses, the bulk metallicities of Kepler-1514b and Kepler-1704b translate to $\sim 100$~\Mearth and $\sim 160$~\Mearth respectively in absolute metal content.

\subsection{How massive are the inner super-Earths?}

Our radial velocity data set has relatively sparse sampling, and is therefore not very sensitive to the radial velocity signals of the three inner super-Earths. 
We quantify the expected RV semi-amplitudes for each planet by using the non-parametric mass-radius (M-R) relationship for \emph{Kepler} planets from \cite{Ning2018} to calculate predicted masses for these planets using their measured radii. We use the \texttt{mr-exo} package to obtain posteriors for the masses of the super-Earths using normal distributions for the radii with mean values and distribution widths from Table~\ref{tab:planet} \citep{Ning2018, Kanodia2019}. The predicted planet masses for Kepler-167 b, c, and d are $4.5^{+6.5}_{-2.6}$ \Mearth, $4.4^{+6.3}_{-2.6}$ \Mearth, and $3.6^{+5.2}_{-2.1}$ \Mearth respectively. These uncertainties are dominated by the relatively large measurement errors and correspondingly large intrinsic scatter for planets with measured masses in this size range. For median mass estimates of these three planets, we would expect RV semi-amplitudes of 2.1 m s$^{-1}$, 1.7 m s$^{-1}$, and 1.0 m s$^{-1}$ respectively, which are well below the noise floor of our data. 

\begin{figure}
    \centering
    \includegraphics[width=\linewidth]{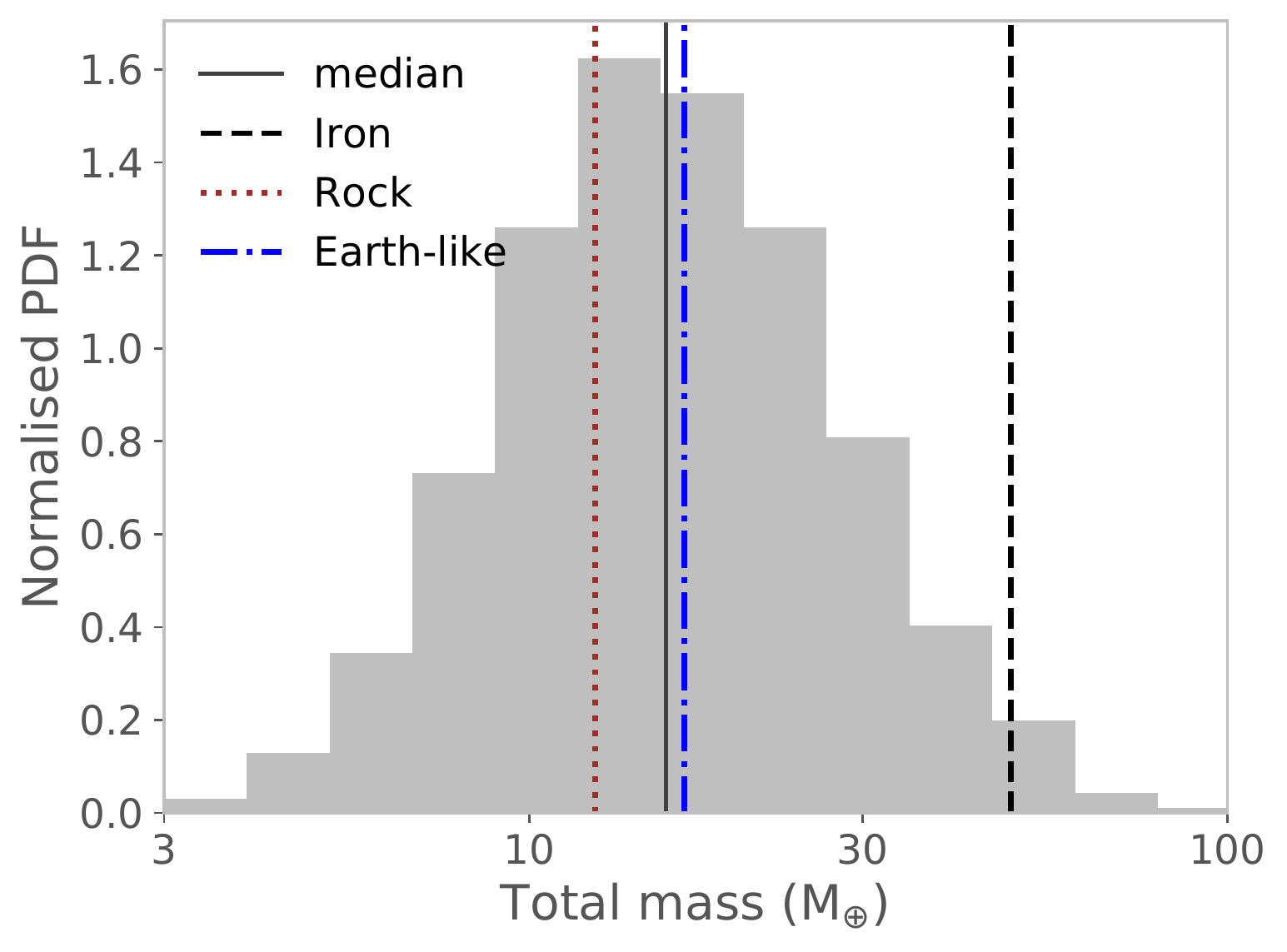}
    \caption{Posterior for the total mass contained in the three inner super-Earths obtained using \texttt{mr-exo} \citep{Kanodia2019}, which utilizes the mass-radius relationship from \cite{Ning2018}. The median of the distribution is shown with a grey line. We also mark the total mass contained in three planets assuming they are pure rock (MgSiO$_3$, brown), iron (black), or Earth-like (blue) using the median radii of these planets from Table 3 and M-R relations from \cite{Zeng2019}.}
    \label{fig:se_mass}
\end{figure}

In order to comment on the solid budget required to form the inner super-Earths, we also need to estimate the total amount of solids contained in these planets. In Figure~\ref{fig:se_mass}, we show the posterior for the total mass contained in the three planets ($15.7^{+11.6}_{-6.5}$ \Mearth) as well as total mass estimates for median planetary radii assuming they are made of pure rock or pure iron \citep{Zeng2019}. The measured radii and orbital periods of these planets place them at or below the location of the radius valley \citep{Fulton2017}. It is therefore unlikely that they host significant hydrogen-rich envelopes \citep{Rogers2015, Owen2017, Ma2021}. The $15$ \Mearth peak of the posterior probability distribution is equivalent to the predicted value for Earth-like rock-iron compositions, and we therefore adopt it as our baseline value for all subsequent calculations. How does this mass compare with the solid mass budget in the inner disk? Since disk density profiles are poorly constrained by observations, we use the Minimum Mass Solar Nebula (MMSN) and Minimum Mass Extrasolar Nebula (MMEN) as baselines to estimate the dust content of the inner disk \citep{Chiang2010, Chiang2013, Dai2020}. The MMSN and MMEN predict $7$ \Mearth and $36$ \Mearth of solids within the orbit of the giant planet respectively. Even in the more optimistic MMEN, the formation of Kepler-167's super-Earths would require dust to be converted to planets with a fairly high efficiency of 40\%. The predicted efficiency of converting dust to super-Earths by either pebble accretion or planetesimal accretion is expected to be $10-20\%$ instead \citep{Drazkowska2016, Liu2018, Ormel2018, Lenz2020}. Moreover, dust in the inner disk is likely to drift into the star on very short timescales. This suggests that the initial solid budget of the inner disk was very likely supplemented by the addition of small solids from regions exterior to Kepler-167e's orbit, which could have migrated inward via radial drift. We explore this scenario in more detail below.

\section{Formation of inner super-earths with outer gas giant companions}
\label{sec:formation}

We use our observational constraints on the properties of the Kepler-167 planets to explore potential formation scenarios for this system. In particular, we are interested in how the presence of a growing giant planet core affects the dust distribution in the disk, since the dust content of the inner disk determines the potential for close-in super-Earth formation. For the giant planet core, we assume that it grows by accreting the marginally coupled `pebbles'. In the pebble accretion paradigm, the growth of the giant planet core depends on the pebble flux through the disk and a threshold pebble flux is typically required to form a sufficiently large core prior to the dispersal of the gas disk \citep[e.g.,][]{Bitsch2019}. We do not consider planetesimal accretion for the formation of the cold giant planet's core \citep[e.g.,][]{Schlecker2020} because it is highly inefficient at the relevant orbital distances unless planetesimals are assumed to be small and turbulent stirring is assumed to be very weak \citep{Johansen2019a}. For the inner super-Earths, both pebble and planetesimal accretion appear to be feasible. However, super-Earths progenitors are likely to dynamically evolve and merge after reaching pebble/planetesimal isolation masses \citep{Dawson2016, Lambrechts2019}, which significantly complicates their formation modeling. We therefore do not model their formation explicitly and instead impose the condition that the amount of solids that reaches the inner disk must be sufficient to form a system of close-in super-Earths (see \S~\ref{sec:form_cj_se}).

In order to understand the formation of the Kepler-167 system, we must therefore first understand the dynamical evolution of solids throughout the disk, which determines the local pebble flux. These pebbles could be directly accreted by the growing protoplanet, or could form planetesimals. The pebble flux is very sensitive to the assumed protoplanetary disk properties such as disk mass, size, metallicity, and turbulence as well as material properties such as the fragmentation velocity of grains \citep[e.g.,][]{Drazkowska2021}. Since our knowledge of these properties is incomplete, we explore a broad parameter space of potential disk models. Although these models are motivated by a desire to explain the origin of the Kepler-167 system, we do not make any star-specific assumptions other than the stellar mass. This means that the models presented here are broadly applicable to all sun-like stars.

\subsection{Protoplanetary disk model}
\label{sec:modelling}

We utilize a simple two-population dust evolution model \citep{Birnstiel2012} as implemented in the publicly available \tpop code\footnote{The original code is available at \url{https://github.com/birnstiel/two-pop-py} and a modified version used in this paper is available at \url{https://github.com/y-chachan/two-pop-py/tree/kepler-167}.} to determine which disks are most conducive for giant planet core formation and to calculate the amount of solids that reaches the inner disk. This model is described in \cite{Birnstiel2012}, which demonstrates that the dust evolution in state of the art numerical simulations is well approximated by splitting the dust population into two groups: one with a spatially and temporally constant size $a_0$ (assumed monomer size $= 0.1 \, \mu$m, corresponding to the mass-weighted average of the grains in the interstellar medium, e.g., \citealp{Laor1993}) and surface density $\Sigma_0$ and the other with size $a_1$ and surface density $\Sigma_1$. The size of the larger grains ($a_1$) is set by growth, drift, and fragmentation and varies as a function of time and location in the disk. This approach allows us to model the dust evolution with a simple advection-diffusion equation:
\begin{equation}
    \frac{\partial \Sigma_{\rm d}}{\partial t} + \frac{1}{r} \frac{\partial}{\partial r} \bigg[r \bigg( \Sigma_{\rm d} \Bar{u} - D_{\rm gas} \Sigma_{\rm g} \frac{\partial}{\partial r} \bigg( \frac{\Sigma_{\rm d}}{\Sigma_{\rm g}} \bigg)  \bigg)  \bigg] = \dot{\Sigma}_{\rm d}.
\end{equation}
where $r$ is the cylindrical distance from the star, $\Sigma_{\rm d} = \Sigma_0 + \Sigma_1$, $\bar{u}$ is the mass weighted velocity of dust grains in the radial direction, $\Sigma_{\rm g}$ is the gas density, $D_{\rm gas}$ is the diffusivity of the gas, and $\dot{\Sigma}_{\rm d}$ is the sink term due to accretion of dust by a growing core. $\dot{\Sigma}_{\rm d}$ is related to the pebble accretion rate, which is discussed in more detail in \S~\ref{sec:pebb_acc}. 

The gas disk is assumed to evolve viscously according to the equations in \cite{Lynden-Bell1974} and its initial surface density profile is derived from the self-similar solution to these equations at time $t = 0$. The radial dependence of the density profile is set by the power law exponent $p$ of viscosity $\nu$, which we fix at unity. The viscosity $\nu$ is calculated as $\alpha_{\rm t} c_s H_{\rm g}$, where \alphat is the Shakura-Sunyaev turbulence parameter \citep{Shakura1973}, $c_s$ is the sound speed, and $H_{\rm g}$ is the gas scale height. The turbulence parameter \alphat and the fragmentation velocity \vfrag of the grains determine the Stokes number of the largest grains, which is given by St$_{\rm frag} = v_{\rm frag}^2 / 3 \alpha_{\rm t} c_s^2$, in fragmentation-limited regions of the disk. Since we are primarily interested in understanding the effect of giant planet core formation on the dust flux in the inner disk, we fix \alphat = 10$^{-3}$ and \vfrag = 10 m s$^{-1}$ to reduce the number of free parameters. Although the values of \vfrag and \alphat are both uncertain, these values are reasonably well supported by the literature \citep[e.g.,][]{Pinte2016, Flaherty2017, Gundlach2015}. This combination of values also ensures that giant planet cores can reach the pebble isolation mass prior to the dispersal of the gas disk. If we wished to form cores in disks with lower values of \vfrag, it would require a correspondingly lower \alphat \citep[e.g.,][]{Venturini2020}.

We explore models for gas disks with characteristic radii of $R_{\rm disk}$ $\in \{20, 60, 200\}$ au, masses $M_{\rm disk}$ $\in \{0.05, 0.1, 0.2\} \, M_*$, and metallicities $Z$ $\in \{0.005, 0.01, 0.02\}$. This allows us to quantify how the pebble flux and the growth rate of the giant planet core vary as a function of each disk parameter. Although we vary $M_{\rm disk}$ and $Z$ as separate model parameters, our results are primarily presented in terms of $M_{\rm dust} = Z \, M_{\rm disk}$, as this quantity plays an important role in controlling the outcome of our models. We fix the mass of the central star to $0.75 \, M_{\odot}$, which is representative of Kepler-167. We calculate the temperature profile of the disk by assuming that it is passively irradiated \citep[e.g.,][]{Chiang1997, DAlessio1998}; this is a reasonable approximation at the location of our giant planet progenitors (see \S~\ref{sec:pebb_acc}): 
\begin{equation}
    T (r) = \bigg[ \phi \; T_*^4 \bigg( \frac{R_*}{r} \bigg)^2 + T_0^4 \bigg]^{1/4}
    \label{eq:temp_struct}
\end{equation}
with a flaring angle $\phi = 0.05$, $T_* = 4180$ K, $R_* = 2.126 \, R_{\odot}$, and $T_0 = 7$ K. We obtain the stellar temperature and radius from a MIST model for a $0.75 \, M_{\odot}$ star at an age of 1 Myr. Although the inner regions of the disk where super-Earths might form are likely to be viscously heated, this has no effect the \emph{integrated} pebble flux that reaches the inner disk from the outer disk.

\subsection{Core formation with pebble accretion}
\label{sec:pebb_acc}
To model the formation of a gas giant core, we introduce a seed of mass $0.01 $ \Mearth at time \tseed, which then grows by accreting solids that drift past its location. Although this initial seed mass is somewhat larger than the predicted masses of planetesimals formed by the streaming instability, this choice allows us to circumvent potential complications related to the unknown initial planetesimal distribution and early growth rates of small planetesimals \citep{Johansen2015, Simon2016}. We vary \tseed $\in \{10^4, 10^5, 10^6\}$ yrs to study the effect of this assumption on the timescale over which the seed reaches the isolation mass. Our chosen \tseed values are motivated by a desire to span a wide range for the time at which a lunar mass seed might form in the disk. We model the growth of cores with final masses of $\in \{10, 15, 20\}$ \Mearth. We do not consider more massive cores even though our model fits indicate that Kepler-167e contains $66 \pm 19$ \Mearth of metals (see \S\ref{sec:k167_bulk_metal}) because the formation timescales for such cores become prohibitively long unless the disks are extremely massive. We adopt the expression for \Miso from \cite{Lambrechts2014a}:
\begin{equation}
    M_{\rm iso} = 20 \bigg(\frac{H_{\rm g}/r}{0.05} \bigg)^{3} \bigg( \frac{M_*}{{\rm M}_{\odot}} \bigg) {\rm M}_{\oplus}.
    \label{eq:iso_mass}
\end{equation}
Although there are updated expressions for \Miso that account for its dependence on other properties of the disk \citep{Ataiee2018, Bitsch2018}, they are in reasonably good agreement with the simpler expression we adopt and depend on parameters we keep fixed in our work (e.g., turbulence, pressure gradient). We determine the location of our seeds implicitly via this expression for the pebble isolation mass. Our initial seeds are therefore placed at 3.2~au, 5.6~au, and 8.2~au respectively, in order to produce cores of $10 $ \Mearth, $15 $ \Mearth, and $20 $ \Mearth. Although these seeds might alternatively have started farther out and then migrated inward as they grew, the predicted migration rates are uncertain and depend sensitively on local disk properties \citep[e.g.,][]{Rafikov2002, Li2009, Paardekooper2010, Benitez-Llambday2015}. We therefore elect to keep the location of each seed fixed in our models. This simplifies our calculations and gives us a conservative lower limit on the dust mass that reaches the inner disk, as an inward-migrating core that starts further out will reach the same isolation mass later, thus increasing the amount of solids that reaches the inner disk. We calculate the growth rate of the core as the accretion rate of dust of size $a_1$ (larger grain population):

\begin{equation}
    \dot{M} = f_{\rm 3D} \dot{M}_{\rm 2D} 
    \label{eq:mdot}
\end{equation}
where $\dot{M}_{\rm 2D}$ is the standard 2D pebble accretion rate in the Hill (shear) regime \citep{Lambrechts2014}:

\begin{equation}
    \dot{M}_{\rm 2D} = 2 \bigg( \frac{\rm min(St_{1}, 0.1)}{0.1} \bigg)^{2/3} R_{\rm Hill}^2 \Omega_{\rm K} \Sigma_{1}.
\end{equation}
$R_{\rm Hill}$ is the core's Hill radius, $\Omega_{\rm K}$ is the orbital frequency at the core's location, and St$_1$ is the Stokes number of grains of size $a_1$. Our assumption of accretion in the Hill regime is justified because the transition mass between the Bondi and the Hill regimes \citep{Johansen2017} is smaller than our adopted seed mass ($0.01$ M$_{\oplus}$) through most of our simulated domain ($< 8$ au, our outermost seed is located only slightly further out at 8.2 au). The factor $f_{\rm 3D}$ in Equation~\ref{eq:mdot} accounts for the effect of the relative magnitudes of the dust scale height and the core's $R_{\rm Hill}$ on the accretion rate \citep{Morbidelli2015}:
\begin{equation}
    f_{\rm 3D} = {\rm min} \bigg(1, \frac{1}{2} \sqrt{\frac{\pi}{2}} \bigg( \frac{\rm St_{1}}{0.1} \bigg)^{1/3}  \frac{R_{\rm Hill}}{H_{\rm d,1}} \bigg)
    \label{eq:f3d}
\end{equation}
where ${H_{\rm d,1}} = H_{\rm g} \sqrt{\alpha_{\rm t} / (\alpha_{\rm t} + {\rm St_{1}} )}$ is the scale height of the large dust grains \citep{Dubrulle1995}. We allow the core to grow until it reaches \Miso, and record the corresponding time \tiso. We assume that this event effectively truncates the flow of solids to the inner disk.

\subsection{Model results}
\subsubsection{Which disks form giant planets?}
\label{sec:gp_form}

\begin{figure}
    \centering
    \includegraphics[width=\linewidth]{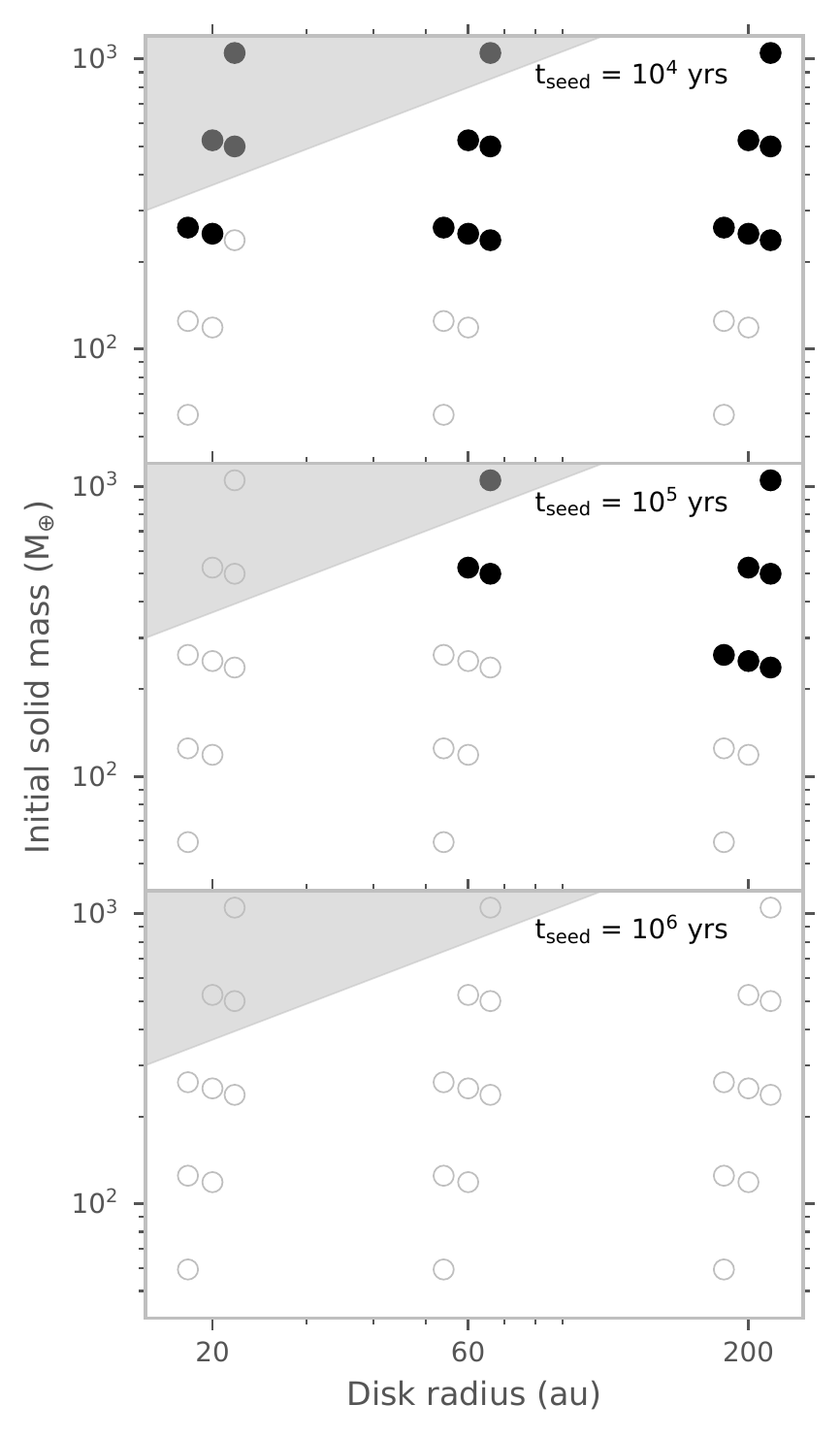}
    \caption{Filled (empty) circles mark disk models for which a $0.01 $ \Mearth seed at 5.6 au does (does not) reach an isolation mass of 15 \Mearth core. We show results from models with different initial solid masses $M_{\rm dust}$ (y-axis, product of $M_{\rm disk}$ and $Z$), disk sizes (x-axis), and three different \tseed. The grey shading indicates combinations of solid mass and disk size that are unlikely to exist in nature. For each combination of disk radius and initial solid mass, we use small offsets to show results for different $M_{\rm disk}$ $\in \{0.05, 0.1, 0.2\} \, M_*$ and $Z$ $\in \{0.005, 0.01, 0.02\}$, with $M_{\rm disk}$ increasing in the horizontal direction and $Z$ increasing in the vertical direction.}
    \label{fig:inital_mdust_vs_rc}
\end{figure}

We determine which of our models are able to successfully form gas giant planets by checking to see which cores reach the isolation mass prior to the dispersal of the gas disk, which we assume occurs at 10 Myr. Figure~\ref{fig:inital_mdust_vs_rc} shows results from the full grid of disk models for $M_{\rm iso} = 15$ \Mearth. Disks where the core reaches the isolation mass are marked with filled circles, while those where it does not are shown as open circles. The importance of the disk's initial solid reservoir is readily apparent \citep[e.g.,][]{Bitsch2019, Schlecker2020}. If the seed of the giant planet core is introduced early (\tseed = $10^4$ yrs, top panel), its ability to reach the isolation mass is determined by the initial solid mass for all but the most compact disk models.

If the seed is introduced later (\tseed = $10^5$ yrs, middle panel), it can only reach the pebble isolation mass if it is located in a relatively large disk. This is because a larger fraction of solids are distributed further out in larger disks, and it takes correspondingly longer for the solids to drain onto the star. This means that seeds that are introduced later can still accrete enough solids to reach the isolation mass (see also \citealp{Johansen2019}). For a fixed solid mass reservoir and disk size, we find that the influence of the total disk mass and dust-to-gas ratio, which we only vary by a factor of a few in these models, is relatively weak. It is the product of disk gas mass and dust-to-gas ratio that really matters.

When the core seed is introduced very late (\tseed $ = 10^6$ yrs, bottom panel), it does not reach the isolation mass in any of the models in our grid. We conclude that \tseed $\lesssim 10^5$ yrs is a requirement for lunar mass seeds to turn into giant planet cores in the framework considered here. We find that seeds introduced at $\lesssim 10^5$ yrs typically reach isolation mass by $\lesssim 1$ Myr (Figure~\ref{fig:Miso_tseed_vary}). This is consistent with the detection of gaps in protoplanetary disks as young as a few Myr old \citep[e.g.,][]{Andrews2018, Long2018}, which are likely opened by planets that are already larger than the pebble isolation mass by this time.

\subsubsection{Which systems with outer gas giant planets also form inner super-Earths?}
\label{sec:form_cj_se}

\begin{figure}
    \centering
    \includegraphics[width=\linewidth]{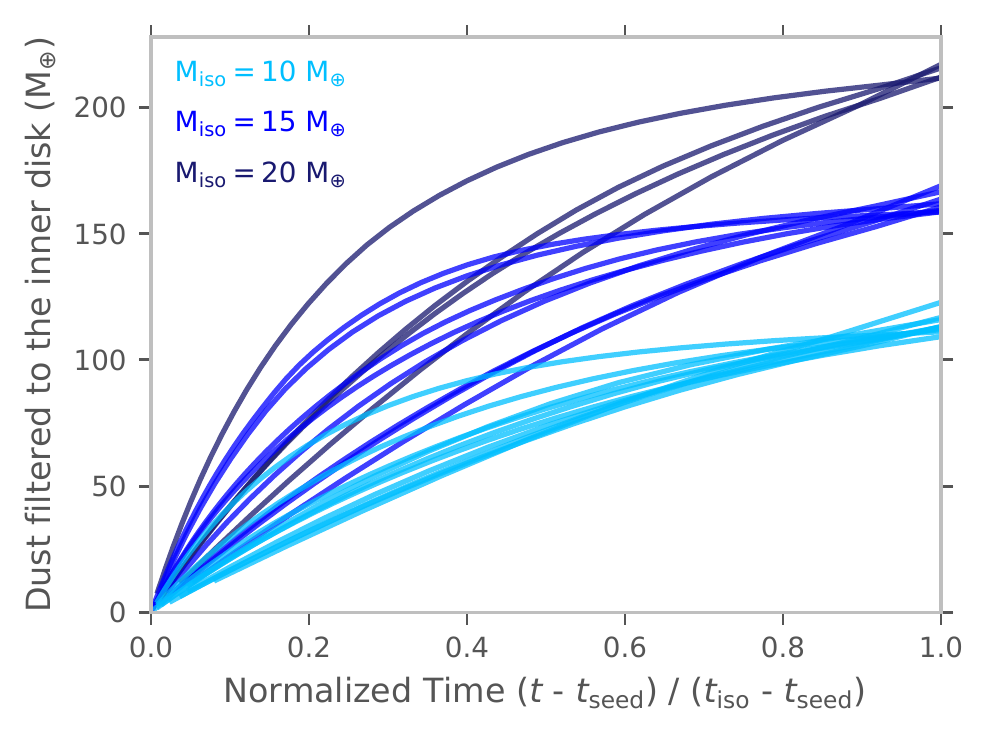}
    \caption{The dust mass that filters through to the inner disk between \tseed (the time at which a lunar mass seed is introduced) and \tiso (when core reaches \Miso) for different \Miso and a fixed \tseed $= 10^5$ years. All disk models in which a lunar mass seed reaches \Miso are shown. The filtered dust mass is primarily a function of \Miso and does not depend strongly on the assumed disk properties.}
    \label{fig:filtered_mdust_vs_normed_time}
\end{figure}

When the giant planet core reaches the pebble isolation mass, the solid reservoir available for planet formation interior to the giant planet's orbit becomes effectively isolated from the outer disk. Although there may still be a modest inward flux of dust across the gap opened by the planet, it is expected to be a few orders of magnitude smaller than the flux in a smooth disk \citep[e.g.,][]{Morbidelli2012, Lambrechts2014a, Drazkowska2019}. This means that the mass budget for planet formation in the inner disk is simply the sum of the initial solid reservoir and the cumulative amount of solids delivered from the outer disk before the gas giant core reaches the isolation mass. The initial solid reservoir in the inner disk is typically negligible compared to the flux from the outer disk for all but the smallest disks.

\begin{figure*}
    \centering
    \includegraphics[width=0.49\linewidth]{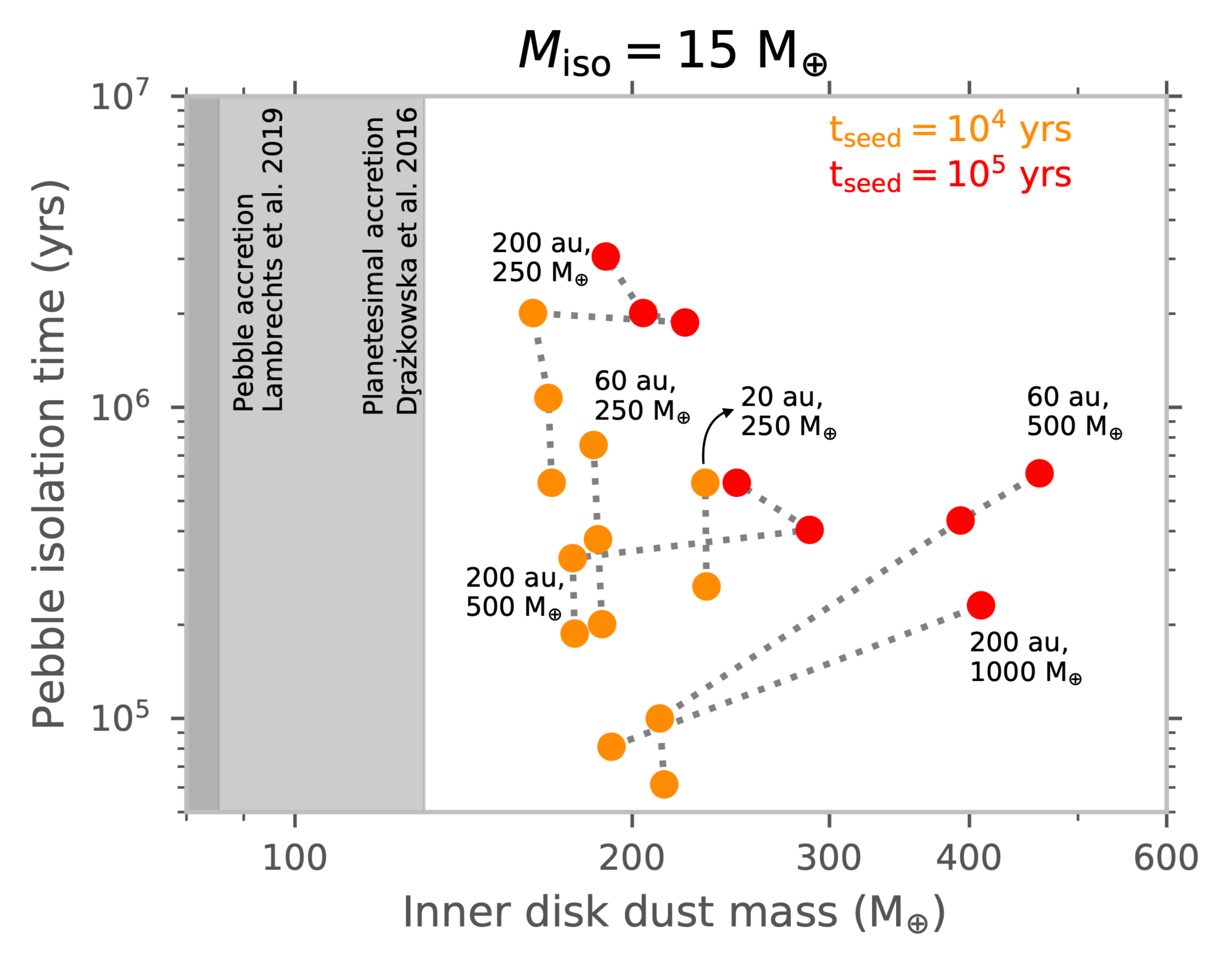}
    \includegraphics[width=0.49\linewidth]{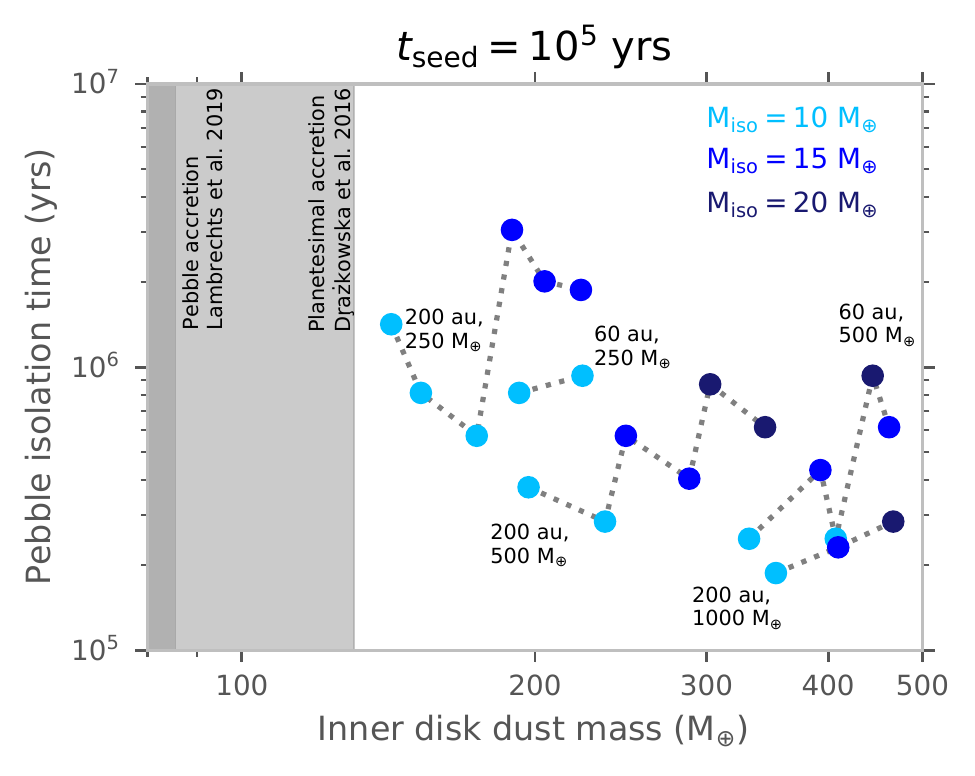}
    \caption{Pebble isolation time vs total solid mass available in the inner disk for all disk models in which a $0.01 $ \Mearth seed reaches \Miso (i.e., those containing an outer gas giant). In the left panel, we vary \tseed and fix \Miso to $15$ \Mearth. In the right panel, \tseed is fixed to $10^5$ yrs and \Miso is varied. The initial seeds are placed at 3.2~au, 5.6~au, and 8.2~au to produce cores of $10 $ \Mearth, $15 $ \Mearth, and $20 $ \Mearth respectively. We mark the estimated dust masses that are required for super-Earth formation in the pebble \citep{Lambrechts2019} and planetesimal \citep{Drazkowska2016} accretion paradigms using shaded regions. Models with the same disk size and initial solid mass are connected via dotted lines.}
    \label{fig:Miso_tseed_vary}
\end{figure*}

The next question that arises is: how much solid mass needs to be delivered to the inner disk for close-in super-Earths to form? The required mass depends on how super-Earths are formed and is likely to be model dependent. We adopt two illustrative limits from the pebble and planetesimal accretion paradigms that provide us with useful estimates of the dust mass needed to form super-Earths. Assuming super-Earths form by accretion of `dry' pebbles onto lunar mass seeds, \cite{Lambrechts2019} show that an integrated pebble flux $\gtrsim 190 $ \Mearth is necessary to form systems of super-Earths with masses and orbital architectures comparable to those observed by \emph{Kepler}. In their models, a factor of 2 increase in the pebble flux (from $100 $ \Mearth Myr$^{-1}$ to $200 $ \Mearth Myr$^{-1}$) changes the final outcome from widely-spaced terrestrial planets to compact systems of close-in super-Earths. Accounting for 50\% mass loss for pebbles across the water snowline, a higher pebble accretion efficiency of our prescription, and the inverse dependence of the accretion efficiency on stellar mass (see Appendix~\ref{sec:acc_eff}), we modify this threshold to $190 / 0.5 \times 3/10 \times 0.75 \sim 86$ \Mearth\footnote{We note that \cite{Lambrechts2019} quantified the pebble mass required to form super-Earths after lunar mass seeds had formed already. This pebble mass threshold does not include the pebble mass required to form the seeds in the first place.}. This modification is likely to be imperfect because the pebble mass threshold is sensitive to various time- and space-dependent quantities. Nonetheless, the key point is to compare the pebble mass threshold for super-Earth formation and the accretion efficiency of a cold giant planet core in the same framework, which we endeavor to do in our study. A less (more) efficient pebble accretion prescription would increase (decrease) the threshold mass for super-Earth formation but it would also increase (decrease) the pebble mass that filters past the cold giant planet core and reaches the inner disk.

Alternatively, super-Earths might form by planetesimal accretion. To quantify the dust mass needed to form super-Earths in this paradigm, we need to know the efficiencies with which i) dust is converted into planetesimals and ii) planetesimals are converted into super-Earths. Unfortunately, quantifying the efficiency of converting dust into planetesimals is quite challenging and there are few estimates in the literature. Here, we use the results of \cite{Drazkowska2016} who use global dust evolution models coupled with planetesimal formation by the streaming instability to show that $\sim 23\%$ of their dust mass is converted into planetesimals. The planetesimals in \cite{Drazkowska2016} form interior to their location of the water snowline so we additionally account for 50\% mass loss of the pebbles that form these planetesimals. Assuming that planetesimals are converted into planets by mutual collision and growth with a 100\% efficiency (commonly assumed in this paradigm), the combined mass $\sim 15$ \Mearth of the Kepler-167 super-Earths translates to $\sim 15 / 0.23 / 0.5 \sim 130$ \Mearth of solids required for formation by planetesimal accretion.

Figure~\ref{fig:filtered_mdust_vs_normed_time} shows the dust mass delivered to the inner disk between \tseed and \tiso, i.e. while the core is growing, for different disks in which the seed reaches \Miso. We find that this integrated dust mass is primarily a function of \Miso and depends only weakly on disk properties (disk size and dust mass). It also has a weak dependence on \tseed itself, i.e. when the seed the introduced (not shown in the plot). This is because the amount of dust mass filtered through to the inner disk while the core is forming depends on the pebble accretion efficiency $\epsilon$ (see Appendix~\ref{sec:acc_eff}) and most of the parameters that affect $\epsilon$ are constant for our disk models (e.g., \alphat, temperature structure). A larger \Miso results in the delivery of a larger amount of solids to the inner disk because seeds take longer to reach a larger \Miso. This is because the larger \Miso places the seed at a larger orbital separation where the disk aspect ratio is higher and pebble accretion efficiency is lower. The amount of dust mass delivered to the inner disk between \tseed and \tiso is $\sim 10 \times$ \Miso, which implies that cumulative $\epsilon \sim 10\%$ for our disk models. With such efficiencies, the dust mass delivered between \tseed and \tiso alone is enough to form inner super-Earths via planetesimal or pebble accretion for \Miso $\gtrsim 10$ \Mearth. This inflowing material is augmented by the initial dust located interior to the giant planet's orbit, as well as the dust mass delivered before \tseed.

Figure~\ref{fig:Miso_tseed_vary} shows the pebble isolation time for the outer giant companion and the corresponding total solid mass available in the inner disk for a range of \tseed and \Miso = $15 $ \Mearth (left panel) and for a fixed \tseed = $10^5$ yrs with varying \Miso (right panel). For a fixed \Miso (left panel), the amount of solids that reaches the inner disk generally increases with \tseed. This is primarily because of the increase in dust mass supplied to the inner disk by radial drift before \tseed, and not because of differences in the dust mass delivered between \tseed and \tiso. We note that for a given \tseed, models with different disk gas masses and dust-to-gas ratios but the same total dust mass have fairly different \tiso, even though they allow roughly the same mass of solids to reach the inner disk. Although we consider \tseed values as low as $10^4$ yrs, we find that there are many potential disk models with enough solids to form super-Earths. This implies that no temporal fine-tuning in the giant planet core's formation is necessary in order to enable the formation of inner super-Earths. For a fixed \tseed ($10^5$ yrs in the right panel of Figure~\ref{fig:Miso_tseed_vary}), a larger \Miso results in the availability of a larger amount of solids for super-Earth formation in the inner disk. In this panel, models with the same \Miso but different disk properties have different total solid mass available in the inner disk due to the disk dependent contribution of radial drift before \tseed.

Variations in disk properties, \tseed, and \Miso lead to a large range in the dust mass available for planet formation in the inner disk. For the most massive disks, the dust mass supplied to the inner disk can significantly exceed the threshold dust mass required to form super-Earths. This might lead to the formation of inner planets with higher masses. For example, \cite{Lambrechts2019} show that increasing the total available pebble mass from 190 \Mearth to 340 \Mearth moved the range of planet masses produced from $2-20$ \Mearth to $5-30$ \Mearth and increased the mean mass of the planets that form by a factor of 2. The most massive disks may therefore also allow for the formation of massive sub-Neptunes and Neptunes interior to a cold giant planet's orbit. This could possibly explain how planetary systems with such architectures emerge (e.g., HAT-P-11, \citealp{Yee2018}; HD 47186, \citealp{Bouchy2009}).

\subsubsection{Constraints on Kepler-167's protoplanetary disk properties}\label{sec:disks}

We can use our grid of disk models to constrain the properties of Kepler-167's protoplanetary disk. We know that: 1) Kepler-167e's core must reach the isolation mass well before the dissipation of the disk, 2) enough solids must be delivered to the inner disk prior to this point to allow for super-Earth formation, and 3) after Kepler-167e's core reaches the isolation mass, there must be enough solids still present beyond its orbit to account for its remaining bulk metal content ($\sim 66$ M$_{\oplus} - $ \Miso). By taking these three conditions into account, we can place a lower limit on the initial dust mass of the disk as a function of disk size. For condition 1, we adopt a stricter limit of 1 Myr rather than our prior 10 Myr for \tiso as we know that Kepler-167e had enough time to accrete a relatively massive (i.e., Jupiter-like) gaseous envelope. This limit is also in better agreement with observational constraints on average disk lifetimes for isolated sun-like stars, which are around 3 Myr \citep[e.g.,][]{Mamajek2009, Williams2011}. Our new upper limit on \tiso excludes scenarios with \tseed of $10^6$, leaving us with a choice between \tseed of $10^4$ and $10^5$ yrs. However, for this exercise we only use \tseed = $10^5$ yrs as $10^4$ yrs is likely too early for a lunar mass seed to form \citep{Lambrechts2012, Visser2016, Johansen2017}. For condition 2, we adopt the higher limit of 130 \Mearth for the dust mass required for super-Earth formation that is imposed by planetesimal accretion \citep{Drazkowska2016}.

Given the sensitivity of gas accretion rates to core mass \citep[e.g.,][]{Lee2019}, we also limit our models to \Miso of 15 and 20 \Mearth, which are more likely to produce a Jovian-mass planet. We note that our results are not qualitatively different for \Miso of 10~\Mearth. Although this requires the giant planet core to accrete additional solids after reaching \Miso in order to match the bulk metal content of Kepler-167e ($\sim 66$ \Mearth), this is a more plausible scenario than models in which the pebble isolation mass is set to 66 \Mearth. Cores of this size can only form in the most massive and largest disk in our grid ($\sim 1000$ \Mearth solids, disk size of 200 au). Since Kepler-167e's bulk metal content is typical for planets in its mass range (see \S~\ref{sec:k167_bulk_metal} and \citealp{Thorngren2019}), it seems unlikely that all of these giant planets formed with such a large \Miso. Planets that have reached the pebble isolation mass may continue to accrete solids in the form of planetesimals \citep[as suggested for Jupiter,][]{Alibert2018} or they might accrete the pebbles that grow from the dust present the circumplanetary disks \citep[e.g.,][]{Drazkowska2018}. Indeed, \cite{Thorngren2016} argue that the late stage accretion of planetesimals is needed in order to explain the mass-metallicity relation observed for extrasolar giant planets. We do not model this process explicitly here but simply require that the remaining solid content at orbital separations beyond the giant planet's core is equal to or greater than 66 \Mearth - \Miso at the time when the core reaches the isolation mass.

The initial dust mass of the disk is a product of the disk dust-to-gas ratio and disk gas mass. In \S~\ref{sec:form_cj_se}, we showed that varying the disk dust-to-gas ratio and gas mass while keeping the total dust mass constant does not affect the solid mass that reaches the inner disk. We therefore reduce the dimensionality of our original grid by fixing the dust-to-gas ratio to 0.015, taking the median stellar [Fe/H] = 0.02 and assuming solar [Fe/H] = 0.014 \citep{Asplund2009}. We are left with a grid in which we vary \Miso, \tseed, disk size, and disk gas mass. We then identify the subset of models in this grid that fulfill the three conditions listed above. In practice, we find that the second condition ($\gtrsim 130$ \Mearth supplied to the inner disk) is automatically met when the first and third conditions are satisfied.

\begin{figure}
    \centering
    \includegraphics[width=\linewidth]{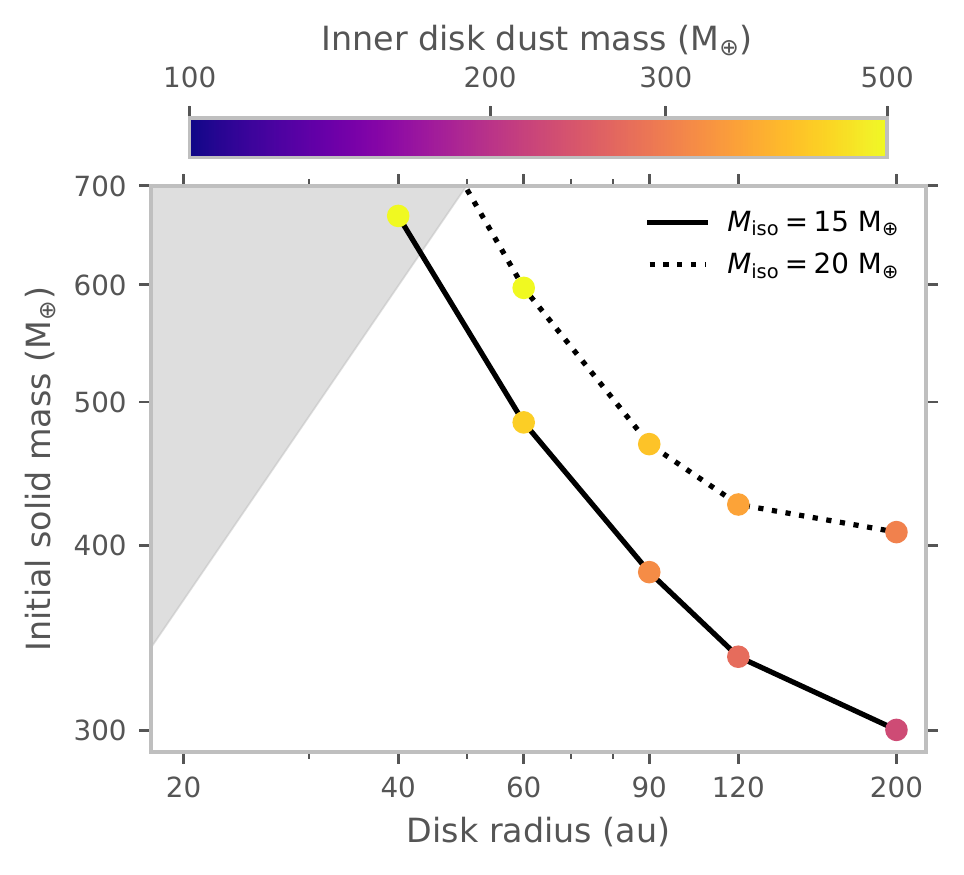}
    \caption{The initial solid mass and size of protoplanetary disks that can produce the Kepler-167 planetary system assuming \tseed = $10^5$ yrs. The color of the points indicates the total amount of solids that is available in the inner disk. We find that $\geq 165 $ \Mearth of solids reach the inner disk for all our models and thus they all exceed the super-Earth formation threshold. We grey out the region corresponding to small disks with very large solid masses, as these disks are unlikely to exist in practice.}
    \label{fig:k_167_disk}
\end{figure}

Figure~\ref{fig:k_167_disk} shows the resulting constraints on the size and initial solid mass of Kepler-167's protoplanetary disk. We find that we require an initial solid mass larger than $\sim300$ \Mearth and a radius larger than 40 au in order to explain this system's present-day properties. Within this range, disks with a larger \Miso require higher initial solid masses in order to form Kepler-167e. For our chosen \tseed $= 10^5$ yrs, the requisite dust mass rises sharply with decreasing disk size. This is primarily driven by the need to have sufficient solid mass beyond the giant planet to explain its bulk metal content (condition 3). Since smaller disks have shorter radial drift timescales and dust rapidly drains out of their outer regions, they need to have larger dust masses to meet this requirement.

\begin{figure}
    \centering
    \includegraphics[width=\linewidth]{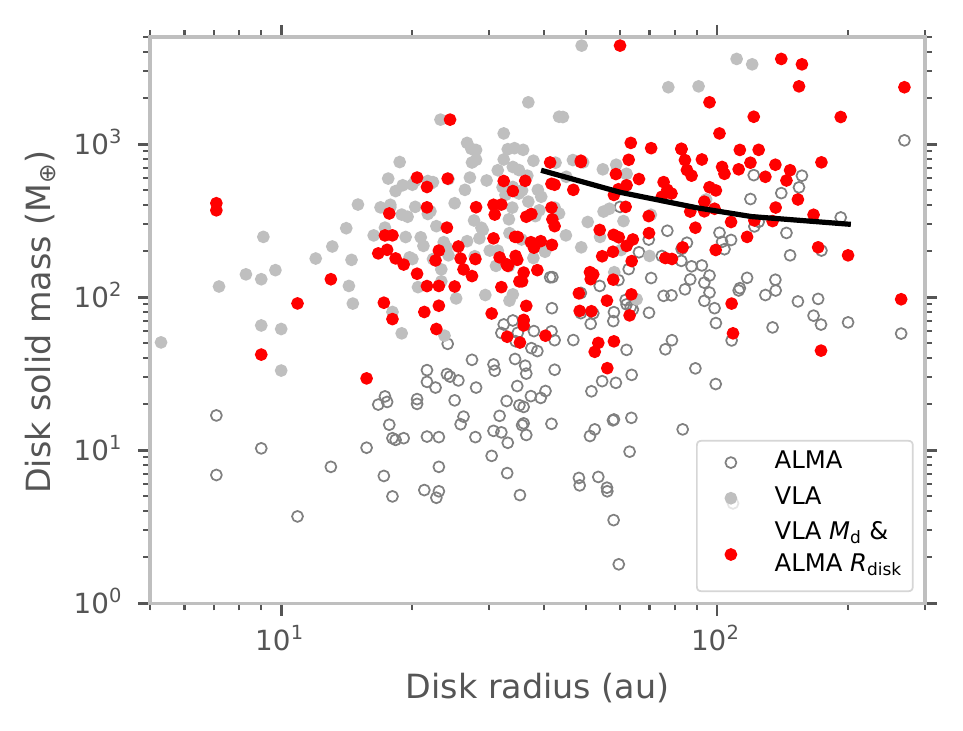}
    \caption{Disk dust mass and radius estimates for Class 0 and I sources in the Orion cluster that are detected both with ALMA (0.87 mm) and VLA (9 mm) \citep{Tobin2020}. We plot the threshold contour above which disks can form systems like Kepler-167 (corresponding to \Miso = 15 M$_{\oplus}$ and \tseed = $10^5$ yrs curve in Figure~\ref{fig:k_167_disk}). Since disks tend to be optically thin in the VLA bandpass, dust mass estimates obtained from these observations are closer to true estimates. However, disk sizes obtained from ALMA are likely to closer to the characteristic disk size that is used in our modelling. We therefore plot dust mass estimates from VLA against disk radii from ALMA in red.}
    \label{fig:disk_demo_k167_const}
\end{figure}

We next consider whether or not these constraints are consistent with results from protoplanetary disk surveys. In Figure~\ref{fig:disk_demo_k167_const}, we plot the ALMA and VLA disk radii and dust masses estimated for Class 0 and I sources in the Orion cluster \citep{Tobin2020} and compare them to the theoretical constraints from our models. Since we are interested in the initial dust mass and size, we exclude Class II disks, which show significant signs of processing, especially for dust mass \citep[e.g.,][]{Tychoniec2020}. If we consider the VLA and ALMA measurements in isolation, we find that very few disks lie above the planet formation threshold contours we have for Kepler-167. However, disk radii estimated from VLA and disk dust masses estimated from ALMA tend to be underestimates. This is evident when we instead plot dust masses derived from VLA against disk radii obtained from ALMA for the same disks (shown in red). Doing so moves the ALMA points up in dust mass and the VLA points to larger radii. Dust masses derived from VLA are likely closer to true values as disks are much more likely to be optically thin at 9 mm than at 0.87 mm. Similarly, since disks tend to appear smaller in continuum emission at larger wavelengths, the radii estimated from ALMA are likely to be closer to the characteristic disk radii that we have in our models. When we combine dust masses from VLA with disk radii from ALMA, we find that a substantial fraction of the disks meet the threshold dust mass and disk size necessary for the formation of the Kepler-167 system.

The formation threshold for the Kepler-167 system is primarily driven by the properties of Kepler-167e, in particular its bulk metal content and the need to form it early. Since Kepler-167e is fairly representative of giant planets beyond several au around FGK stars, we can roughly quantify the fraction of disks around single FGK stars that lie above our formation threshold ($f_{\rm disk}$) and compare it with the corresponding occurrence rate of giant planets \citep[e.g.,][]{Wittenmyer2020, Fulton2021}. Of the 425 disks targeted by ALMA in \cite{Tobin2020}, 45 disks lie above the \Miso = 15 \Mearth threshold. However, this sample is likely to contain both massive and low mass stars that will bias our estimate of $f_{\rm disk}$. Correcting for this contamination as well as the presence of close companions to FGK stars that likely go undetected in \cite{Tobin2020} (see Appendix~\ref{sec:fdisk} for details of this correction), we find that $f_{\rm disk} \approx 14\%$.

We conclude that it is reasonably probable that a star with Kepler-167's mass might host a disk with an initial solid mass and radius that lie above the thresholds indicated by our disk models. If we take the giant planet occurrence rate beyond several au around FGK stars \citep[e.g.,][]{Wittenmyer2020, Fulton2021} as a proxy for the occurrence rate of Kepler-167-like systems, we find that this value is broadly consistent with our estimated massive disk fraction of $10-20\%$. Our models also suggest that there is likely to be a strong correlation between outer gas giants and inner super-Earths, as most disks that met our conditions for giant planet formation also delivered enough material to the inner disk to form short-period super-Earths. We note that massive metal-rich disks are also more likely to form multiple gas giants, which in turn can pump up the eccentricities of the gas giants and destabilize the system of inner super-Earths. That is, post-formation dynamical evolution might reduce the strength of the correlation between inner super-Earths and outer gas giants for metal-rich stars with massive disks. Nonetheless, our results are consistent with observational studies, which find a strong empirical correlation between these two populations \citep{Zhu2018, Bryan2019, Herman2019}.

\section{Conclusions}\label{sec:conclusions}

The fact that close-in super-Earths often accompany cold giant planets provides us with valuable insights into the planet formation process. Systems such as Kepler-167, which can be characterized in detail, serve as an important bridge between observed planetary properties and planet formation models. In this work, we re-fit the \emph{Kepler} photometry in order to derive updated parameters for both the host star and the four transiting planets in the system. We also obtain radial velocity measurements spanning more than three years in order to measure the mass of the outer transiting gas giant, Kepler-167e. We determine that Kepler-167e is a true Jupiter analog with a mass of $1.01^{+0.16}_{-0.15}$ M$_{\rm J}$. Its mass and semimajor axis are typical of gas giant planets detected by radial velocity surveys, but it appears to have a relatively low orbital eccentricity ($3\sigma$ upper limit of 0.29). 

We fit Kepler-167e's measured mass and radius using a giant planet evolution model and find that this planet is more metal-rich than Jupiter, with a bulk metallicity of $0.21 \pm 0.05$. This translates to an impressive $66^{+20}_{-18}$ \Mearth of metals in its interior. Although our RV data are not precise enough to place any constraints on the masses of the inner super-Earths, we use the non-parametric M-R relationship from \cite{Ning2018}, to estimate that the three planets are predicted to contain $15.7^{+11.6}_{-6.5}$ \Mearth in total. Dust in the inner disk drifts into the star on very short timescales and converting the local dust content into planets requires a rather high formation efficiency. It therefore seems unlikely that these planets could have formed from the material initially located inside Kepler-167e's orbit, and instead the dust budget must have been supplemented by the migration of solids from the outer disk. Taken together, these two quantities constrain the initial solid budget of Kepler-167's protoplanetary disk.

We quantify the conditions required to form the Kepler-167 system by exploring a simple grid of protoplanetary disks models spanning a range of initial solid masses and disk radii. We find that giant planets like Kepler-167e preferentially form in fairly massive (in dust content) and large disks, in good agreement with results from previous studies \citep[e.g.,][]{Bitsch2019, Schlecker2020}. Our models assume that once the giant planet reaches the isolation mass, the flow of solids to the inner disk is effectively truncated. Despite this constraint, we find that most disks that form outer gas giants are nonetheless able to supply enough solids to the inner disk to also form super-Earths in both the pebble \citep{Lambrechts2019} and planetesimal \citep{Drazkowska2016} accretion paradigms. This remains true regardless of the time at which the giant planet seed is introduced, and we find consistent results across a range of different isolation masses for the giant planet core.

When we incorporate the additional constraint provided by Kepler-167e's bulk metallicity, we find that we require disks that contains $\gtrsim 300$ \Mearth of solids and are $\gtrsim 40$ au in size to form this planet. We compare these constraints with the observed properties of Class 0 and I disks in the Orion cluster as measured by ALMA and the VLA \citep{Tobin2020}. We find that $10-20\%$ of FGK stars should have disks with masses and radii large enough to form the Kepler-167 system, even after accounting for contamination from a range of stellar populations. This strengthens the plausibility of our constraints on Kepler-167's protoplanetary disk properties.

Further efforts to characterize the Kepler-167 system will enhance our understanding of the origin of its planetary configuration. In particular, there is a pressing need for mass measurements of the inner planets. Our ignorance of the super-Earth planet masses hinders our ability to estimate the accretion efficiency of pebbles and/or planetesimals. Future observations with next generation instruments such as the Keck Planet Finder \citep[KPF,][]{Gibson2016} will allow us to measure the masses of the super-Earths and put our formation scenario on a firmer footing. RV semi-amplitudes corresponding to Earth-like planetary composition for Kepler-167 b and c are expected to be accessible with KPF. Additionally, characterization of Kepler-167e's atmospheric composition would be a useful probe of its formation history and location. However, Kepler-167 is a faint star and Kepler-167e has a high surface gravity, a cold atmosphere, and is a good candidate for photochemical hazes. We also cannot stack multiple transits or eclipses because it transits so infrequently. Taken together, these factors mean that it does not appear to be a good target for atmospheric characterization with JWST.

Although it is outside the scope of this study, we note that further characterization of Class 0 and I disks would be particularly valuable for bridging the gap between disk properties and planet formation. It is fortuitous that we can obtain observational constraints on disk dust masses and radii, as these two quantities have a significant impact on planet formation. There is a growing consensus in the field that planet formation starts earlier than the Class II stage \citep[e.g.,][]{Tychoniec2020, Segura-Cox2020}, and it would therefore be particularly useful to carry out additional comprehensive surveys targeting other young star forming regions with a significant population of Class 0 and I disks. This would allow us to more accurately assess the distribution of disk properties at early times, which we can use to make predictions for giant planet occurrence rates. On the modelling end, our Class II disk model is unlikely to be appropriate for the early stages of disk evolution and therefore more accurate models are needed for these initial epochs. A deeper understanding of the connection between the collapse of protostellar cores and the initial properties of protoplanetary disks, such as disk sizes and the timescale over which dust and gas are delivered, would also help to better elucidate the environment in which planets first begin to form \citep[see][for recent attempts in this direction]{Lebreuilly2020, Lebreuilly2021, Lee2021, Schib2021}.

\section*{Acknowledgements}

We are grateful to the referee for a timely, thoughtful, and instructive report that helped improve this paper. The authors would like to express their gratitude to Dan Foreman-Mackey for assistance with the \texttt{exoplanet} package. Y. C. is grateful to Michael Greklek-McKeon, Michiel Lambrechts, Eve Lee, Lee Rosenthal, and Shreyas Vissapragada for helpful discussions. The authors thank all of the observers in the California Planet Search team for their many hours of hard work. H. K. acknowledges support from NSF CAREER grant 1555095. P. D. is supported by a National Science Foundation (NSF) Astronomy and Astrophysics Postdoctoral Fellowship under award AST-1903811.

This research has made use of the NASA Exoplanet Archive, which is operated by the California Institute of Technology, under contract with the National Aeronautics and Space Administration under the Exoplanet Exploration Program. Some of the data presented in this paper were obtained from the Mikulski Archive for Space Telescopes (MAST) at the Space Telescope Science Institute. The specific observations analyzed can be accessed via \dataset[DOI]{https://doi.org/10.17909/T9059R}. Funding for the \emph{Kepler} mission is provided by the NASA Science Mission Directorate. STScI is operated by the Association of Universities for Research in Astronomy, Inc., under NASA contract NAS 5–26555. 

Some of the data presented herein were obtained at the W. M. Keck Observatory, which is operated as a scientific partnership among the California Institute of Technology, the University of California, and NASA. The Observatory was made possible by the generous financial support of the W. M. Keck Foundation. Finally, the authors recognize and acknowledge the cultural role and reverence that the summit of Maunakea has within the indigenous Hawaiian community. We are deeply grateful to have the opportunity to conduct observations from this mountain.

\facilities{Keck:I (HIRES), Kepler}\\

\software{ \texttt{EXOFASTv2} \citep{Eastman2013,Eastman2019}, 
                \texttt{lightkurve} \citep{lightkurve2018}, 
                \texttt{SpecMatch} \citep{Petigura2015,Petigura2017b}, 
                \texttt{SpecMatch--Emp} \citep{Yee2017}, 
                \texttt{exoplanet} \citep{ForemanMackey2021},
                \texttt{pymc3} \citep{pymc3},
                \texttt{theano} \citep{theano},
                \texttt{starry} \citep{Luger2018,Agol2020}
                \texttt{celerite} \citep{ForemanMackey2017,ForemanMackey2018}
                \texttt{astropy} \citep{astropy2018}
                \texttt{twopoppy} \citep{Birnstiel2012}}

\appendix
\section{Pebble accretion efficiency}
\label{sec:acc_eff}
The pebble flux threshold determined by \cite{Lambrechts2019} for super-Earth formation is applicable for a solar mass star. To obtain the integrated pebble flux threshold for super-Earth formation around a less massive star, we need to determine the stellar mass dependence of the pebble accretion efficiency. The pebble accretion efficiency is equal to the pebble accretion rate divided by the radial pebble flux at the growing planet's location. Depending on whether the pebble scale height is smaller or larger than the Hill radius of the planet (2D or 3D regime), this efficiency $\epsilon$ is given by:
\begin{equation}
    \epsilon_{\rm 2D/3D} = \frac{\dot{M}_{\rm 2D/3D}}{2 \pi r v_r \Sigma_d}
\end{equation}
where $\dot{M}_{\rm 3D} = f_{\rm 3D} \dot{M}_{\rm 2D}$ when $f_{\rm 3D} < 1$ (see Equations~\ref{eq:mdot}-\ref{eq:f3d}). In the expression for the radial pebble flux, $r$ is the radial distance from the star, $\Sigma_d$ is the surface density of pebbles (equal to $\Sigma_1$ in our two-population model), and $v_r = 2 \eta r \Omega_{\rm K} {\rm St}$ is the radial drift velocity of the pebbles. Here, $\eta = -0.5 \; {\rm d \, ln}P / {\rm d \, ln}r \; (H_{\rm g} / r)^2$ is a measure of the deviation of gas' orbital velocity from the Keplerian velocity. Following through, we obtain the following expressions for pebble accretion efficiency in the 2D and 3D regimes:
\begin{equation}
    \epsilon_{\rm 2D} = \bigg(\frac{10}{3}\bigg)^{2/3} \frac{1}{2 \pi} \frac{q^{2/3}}{\eta \, {\rm St}^{1/3}} \approx 0.36 \frac{q^{2/3}}{\eta \, {\rm St}^{1/3}}
\end{equation}

\begin{equation}
    \epsilon_{\rm 3D} = \bigg(\frac{5}{6}\bigg) \frac{1}{\sqrt{2 \pi}} \frac{q \, r}{\eta \, H_{\rm d}} \approx 0.33 \frac{q}{\eta \, h_{\rm d}}
\end{equation}
Here, $q = M_{\rm p} / M_*$ is the mass ratio of the growing planet and $h_{\rm d} = H_{\rm d} / r$. The numerical coefficients and the physical dependencies we obtain match those given in Table 2 of \cite{Ormel2018} (listed under \citealp{Morbidelli2015} and \citealp{Lambrechts2014}, which matches our prescription). The numerical coefficients are in reasonable agreement with the values determined from 3D simulations in \cite{Ormel2018} (our $\epsilon_{\rm 2D}$ is 50\% higher and $\epsilon_{\rm 3D}$ is $15\%$ lower, most of the seeds in our models accrete in the 3D so $\epsilon_{\rm 3D}$ is the relevant value). Our expression for $\epsilon_{\rm 3D}$ is larger by a factor of $10/3$ compared to the value obtained by \cite{Lambrechts2019}.

Super-Earths form in the inner disk and accrete in the 3D regime in \cite{Lambrechts2019}, where the pebble accretion efficiency $\epsilon_{\rm 3D} \propto 1 / M_*$. Therefore, accounting for the higher pebble accretion efficiency (factor of $10/3$), the lower stellar mass (0.75 \msun) in our models, and 50\% mass loss due to sublimation of water ice from pebbles, the threshold for super-Earth formation by pebble accretion is roughly $190 \times 3/10 \times 0.75 / 0.5 \sim 86$ \Mearth. We note that this is the integrated pebble flux that must reach the inner disk after the super-Earth seeds have formed. This threshold therefore does not include the pebble mass required to form the seeds in the first place. We have also ignored the stellar mass dependence of other quantities in the expression, most notably the disk aspect ratio and the Stokes number of pebbles in the inner disk.


\section{Disk fraction and formation threshold}
\label{sec:fdisk}
Of the 425 disks targeted by ALMA in \cite{Tobin2020}, 45 disks lie above the \Miso = 15 \Mearth threshold. However, this sample is likely to contain both massive and low mass stars that will bias our estimate of $f_{\rm disk}$. Restricting the observational sample of protostars to a range of masses is notoriously difficult as protostellar masses are highly uncertain for Class 0 and I sources \citep[e.g.,][]{Dunham2014, Fischer2017}. Instead, we correct for contamination from massive and low mass stars using the measured initial mass function (IMF) for the Orion nebula. In order to do so, we need to know how well the ALMA sample captures the IMF, including the threshold stellar mass below which we miss most protostellar objects. Given the wide range ($10^{-2} - 10^3$ L$_{\odot}$) of protostellar luminosities exhibited by the sources in \cite{Tobin2020}, it is likely that we are only missing protostars that reach $\lesssim 0.1-0.2$ M$_{\odot}$ at the end of accretion \citep[e.g.,][]{Offner2011, Hartmann2016}. Adopting the IMF from \cite{DaRio2012} along with the modification suggested by \cite{Krumholz2012} to account for stars more massive than 2 M$_{\odot}$, we find that these stars constitute $\sim 20\%$ of the stellar population. 

Therefore, assuming that the ALMA survey likely samples only the top 80\% of the IMF, stars more massive than 1.4 M$_{\odot}$ should comprise approximately 12.5\% of our sample (53 sources). We expect a substantial fraction of disks above our formation threshold to be hosted by these massive stars. However, nearly $\sim 50\%$ of massive stars have a companion within log $P = 5$ \citep{Moe2017}, which are associated with lower disk masses and/or lifetimes \citep{Kraus2012}. This means that at most half of the massive stellar population should host massive disks in our survey sample. We must also correct for contamination from M stars that lie above the luminosity threshold. We assume that disks around M stars are likely to have lower masses\footnote{This is at least well supported by observations of Class II disks \citep[e.g.,][]{Andrews2013, Ansdell2016}. For Class 0 and I disks, \cite{Tobin2020} observe a weak correlation between disk mass and protostellar luminosity ($M_{\rm dust} \propto L_{\rm bol}^{0.31 \pm 0.05}$ for non-multiple sources, see their Figure 8), where the latter serves as a relatively poor tracer of stellar mass.} and we therefore do not expect a substantial number of these disks to lie above our formation threshold. Using the assumption adopted earlier that we only miss the bottom 20\% of the IMF, we would expect half the stars in our sample to be $< 0.5$ M$_{\odot}$ (212 sources). Finally, we expect $\sim 20\%$ of the remaining FGK stars (32 out of the remaining 160) to have close-in companions ($\lesssim 50$ au, \citealp{Moe2017}) that we cannot detect\footnote{Although \cite{Tobin2020} indicate whether a set of sources are part of a multiple system or not, they can only resolve companions separated by $\gtrsim 40 $ au. Using the position coordinates, \emph{Gaia} distances to the sources, and disk radii provided in \cite{Tobin2020}, we find that there is likely only a single pair of bona fide binary stars in their sample with separation $< 50 $ au.} and we must therefore remove these from the denominator. Performing all these steps, we find $f_{\rm disk} = (45 - 53/2) / (425 - 53 - 212 - 32) \approx 14\%$.

\newpage

\bibliography{manuscript}
\bibliographystyle{apj}

\end{document}